\newcommand{\refeq}[1]{Eq.$\,$(\ref{#1})}

\documentclass[aps,prb,preprint]{revtex4-1}
\usepackage{amsmath}
\usepackage{graphicx}
\usepackage{subcaption}
\usepackage{color}

\begin{document}
\title{A Monte Carlo method for solving the NEGF equations for electron transport}
\author{Lars Musland}
\author{Joakim Bergli}
\affiliation{University of Oslo, Department of Physics, P.O. Box 1048 Blindern, NO-0316 Oslo, Norway}
\author{Espen Flage-Larsen}
\affiliation{SINTEF Materials and Chemistry, P.O. Box 124 Blindern, NO-0314 Oslo, Norway}
\affiliation{University of Stavanger, Department of Mechanical and Structural Engineering and Materials
Science, Ullandhaug, N-4036 Stavanger}
\begin{abstract}
We derive, by introducing restrictions to the lesser self energy, a Monte Carlo scheme that solves the NEGF equations for electron transport. In doing so we formally prove that the Monte Carlo estimator has an expectation value equal to the lead current of the NEGF solution, and we provide a simple test of the Monte Carlo scheme by calculating conductivity in nanowires within Buttiker's approximation of scattering. Good agreement between the Monte Carlo simulations and the alternative approaches are obtained, and we also demonstrate the existence of a regime where the Monte Carlo method is the fastest method. In our tests this regime is to extreme to be of practical use. We discuss various ways in which to speed up our prototype and how it can be extended to include more physics. Although the full applicability range of our assumptions about the lesser self energy remains to be better understood, we argue that they should apply as long as the transport process can be considered stationary.
\end{abstract}
\maketitle

\section{Introduction}

In the study of electron transport, the most commonly applied formalism is the Boltzmann transport equation, which works well in most applications. However, effects that originate in the wave like nature of electrons are difficult to address with the semiclassical Boltzmann equation. In order to include such effects, it is instead preferable to use a transport formalism derived from the Schr\"odinger equation. An example of such a formalism, is the Non-equilibrium Green's functions (NEGF) \cite{jacoboni}, a generalization of many particle perturbation theory to non-equilibrium. Due to reduction in size and increased complexity of combinations of different materials, recent years has brought significant interest in developing NEGF solvers\cite{nemo,tbsim}.

However, a remaining challenge with NEGF is the computational burden. NEGF is formulated in terms of matrices of size $N\times N$, where $N$ is proportional to the spatial system size, and to the number of included states. In addition, these matrices are functions of energy $E$, and in many cases also of a Bloch vector $\boldsymbol{k}$. To obtain acceptable accuracy, the continuous parameters $E$ and $\boldsymbol{k}$ must be discretized in the form of a dense integration grid. Thus, without the introduction of approximations, NEGF calculations  involve the solution of large equation sets.

In fact, similar challenges also apply when solving the Boltzmann equation. The Boltzmann distribution function, without approximations depends both on position and on Bloch vector. Thus, the number of parameters is still proportional to the system size and large integration grid. By introducing the relaxation time approximation, the procedure can be simplified, particularly in the linear response regime. However, in cases where this approximation is not warranted one must instead solve the general Boltzmann equation. A common approach is to utilize Monte Carlo techniques, which involves the explicit simulation of electron motion, with random scattering between $k$-points\cite{lundstrom, jacoboni}.

Such an approach avoids the explicit solution of large equation sets, and the required computation time is instead determined by the ratio between the estimators variance and the required accuracy. Through the use of such Boltzmann Monte Carlo (BMC) calculations, one has at moderate temperatures, obtained impressive agreement between experimental and theoretical values of transport coefficients in simple semiconductors\cite{jacoboni}. Building on this success, is it possible that a similar approach can employed for NEGF? It can be shown that the BMC method formally corresponds to a Monte Carlo evaluation of the Neumann expansion of the Boltzmann equation. This approach is in fact generalizable to several transport formalisms\cite{jacoboni}.

Monte Carlo simulations have been applied to the calculation of Green's functions in the form of the Diagrammatic Monte Carlo (DMC) method. In this method, the Green's function is estimated through Monte Carlo summation of an infinite series of Feynman diagrams\cite{VANHOUCKE201095,PhysRevLett.81.2514,PhysRevLett.99.250201} and is also similar to a Neumann expansion. The DMC thus shares some features with the NEGF approach to be introduced in this work. However, an important difference is that DMC performs a Monte Carlo evaluation also on the self energy. In the case of electrical transport, typical NEGF solvers\cite{nemo,tbsim} assume the self energies to be given explicitly.

The evaluation of Monte Carlo estimators in quantum transport will generally result in opposite signs or phases. In a large sample one can thus expect partial cancellation\cite{jacoboni}, leading to a variance which is greater than the expectation value. The Monte Carlo integral thus converges slowly. In the DMC method this is partially alleviated by performing some of the diagrammatic summations analytically\cite{PhysRevLett.99.250201}.

In this work, we suggest a Monte Carlo scheme based on a Neumann expansion of the lesser Green's function $G^<$ in the NEGF formalism. In the method suggested here, we will assume that the system is stationary. As will be shown, the Neumann expansion can be expressed in terms of positive quantities, and the ``sign problem'' is avoided. However, this is at the cost of increased computational complexity per estimate.

Briefly, the suggested Monte Carlo method can be described as a process where electrons are randomly scattered from state to state. This is similar to the DMC method and to BMC simulations. However, while states in the BMC simulations are described by a combined position in physical space and $\boldsymbol{k}$-space, we instead describe states using wave functions $|\psi\rangle$. These wave functions are found by solving a modified Schr{\"o}dinger equation with a source term $|i\rangle$. The source term is picked randomly among the eigenvectors of the operator $\Lambda(|\psi'\rangle\langle\psi'|)$, where $|\psi'\rangle$ is the previous state, and $\Lambda$ is a function with the property that $\Lambda(G^<)=\Sigma^<$. $G^<$ being the lesser Green's function and $\Sigma^<$ the lesser self energy. This process is described in more detail in section \ref{intuitive_integral}.

As long as sparse matrices can be used to model the Hamiltonian and the scattering operators, the solution of the modified Schr{\"o}edinger equation will be a linearly scaling operation. Thus, since the calculation of $N$ eigenvectors will always scale at least quadratically with the system size $N$, the bottleneck of the method will generally be the diagonalization of $\Lambda(|\psi\rangle\langle\psi|)$. In order for the method to be practical, it is likely that approximation must be applied in order to make the diagonalization tractable. Among such approximations, is B\"uttiker's approximation, where $\Lambda(|\psi\rangle\langle\psi|)$ is diagonal in some basis $\{|i\rangle\}$, independent of $|\psi\rangle$. In this work, all calculations have been performed within this approximation in order to test the validity of the method.

This work is organized as follows: Section \ref{theory} presents the method. Here we prove that the described Monte Carlo scheme solves the NEGF equations. A discussion of the linear limit of the procedure and a description of  B\"uttiker's approximation is also given. In section \ref{implem} we describe the prototype implementation in more detail.

The results of the initial tests are presented in section \ref{results}, where we calculate the conductance of nanowires. In particular, we study how the Monte Carlo computation time required for a fixed accuracy of one percent scales with the length of the wires, and how this compares to iterative and direct schemes of solution. 

Our prototype Monte Carlo solver is implemented in Python. Still, we demonstrate the existence of a regime where the Monte Carlo calculations are faster than the alternative approaches. This regime is however not to be expected to be relevant in practical applications. Thus, in order for the presented Monte Carlo scheme to be useful, steps must be taken to further increase performance. We do expect a significant improvement upon transition to a compiled language. Further suggestions of how it is possible to reduce the computational burden are discussed in section \ref{discussion}, where we also discuss how more general physics can be included. Finally, section \ref{conclusion} summarizes and presents a few conclusions.

\section{Theory}\label{theory}

This section is divided into the the following parts: Section \ref{negf_summary} gives a quick summary of the NEGF formalism, while in section \ref{it_exp_sec} we derive a Neumann expansion of the NEGF equation, which is needed to define the general Monte Carlo integration scheme. This scheme is derived in section \ref{general_integral}, where the most general version of the method is described. This requires that we assume some probability distribution $g$, from which sequences of electron states $|\psi\rangle$ can be drawn. In section \ref{intuitive_integral} we introduce a concrete procedure which specifies the probability distribution $g$. Finally, in section \ref{lin_limit_section} we discuss the linear limit for elastic scattering, and in section \ref{buttiker_section} B\"uttiker's approximation.

\subsection{Summary of the NEGF formalism}\label{negf_summary}

The Non-equilibrium Green's function formalism for electrons takes its simplest form when expressed in terms of two correlation functions $G^r$ and $G^<$. These are respectively referred to as the retarded and lesser Green's functions, and can be expressed as matrices with elements\cite{jacoboni} 
\begin{align}
G^r_{ij}(t,t')=&-\frac{i}{\hbar}\left\langle\right\{a_i(t),a^\dag_j(t')\left\}\right\rangle\theta(t-t'),\\
G^<_{ij}(t,t')=&+\frac{i}{\hbar}\left\langle a^\dag_i(t)a_j(t')\right\rangle.
\end{align}
Here $a_i$ and $a_i^\dag$ are fermionic annihilation and creation operators, operating on electrons in the single particle state $i$. $\theta(t)$ is the unit step function, curly brackets represents the anti-commutator, and the brackets represent the average with respect to the quantum state.

If the Green's functions describe a stationary process, then they will only depend on the difference between the time arguments, so that we may define the Fourier transforms\cite{datta}
\begin{align}
G^r(E)=&\int\mathrm{d}t\,G^r(t,t')e^{iE(t-t')/\hbar},\\
G^<(E)=&\int\mathrm{d}t\,G^<(t,t')e^{iE(t-t')/\hbar}.
\end{align}
In addition, the formalism makes use of the advanced Green's function $G^a$, which can be found through the relation $G^a(E)=G^r(E)^\dag$, as well as the spectral density
\begin{align}
A(E)=i\left(G^r(E)-G^a(E)\right).
\end{align}
The literature operates with multiple conventions regarding the definition of $A(E)$, but in this work we follow the line of Datta\cite{datta}, and define this without additional prefactors. 

In equilibrium there is a simple relationship between the lesser Green's function and the spectral density, namely\cite{jacoboni,datta}
\begin{align}
G^<(E)=iA(E)f(E),
\end{align}
where $f(E)$ is the Fermi distribution. Away from equilibrium however, there is no such simple relation, and accordingly $G^<$ and $G^r$ must be calculated independently. Within diagrammatic perturbation theory, it can be shown\cite{keldysh,khan1987quantum} that for stationary processes this can be done by solving the equations
\begin{align}\label{eq_for_gr}
G^r(E)=&\left(E-H-\Sigma^r(E)\right)^{-1},\\\label{eq_for_gl}
G^<(E)=&G^r(E)\Sigma^<(E)G^a(E),
\end{align}
where $H$ is the single particle Hamiltonian of the electrons, while $\Sigma^r$ and $\Sigma^<$ are respectively the retarded and the lesser self energies. These self energies will in general be functions of $G^r$ and $G^<$, and their form will depend on the particular perturbation used. Examples of such perturbative expressions can be found in the literature\cite{datta,jacoboni,lake1997single}, while a more heuristic model is discussed in section \ref{buttiker_section}.

In a typical electron transport problem, we are considering a device or material which is connected to a set of leads $\{p\}$. The leads are assumed to be internally close to equilibrium, at least compared to the device, so that their electronic occupations are described by Fermi distributions $f_p(E)$. It can then be shown\cite{datta} that the effect these leads have on the device can be accounted for in terms of the self energies
\begin{align}\label{sigmas_comp_r}
\Sigma^r(E)&=\sum_p\Sigma^r_p(E)+\Sigma^r_s(E),\\\label{sigmas_comp_l}
\Sigma^<(E)&=i\sum_p\Gamma_p(E)f_p(E)+\Sigma^<_s(E).
\end{align}
Here $\Sigma^r_s$ and $\Sigma^<_s$ are self energies due to scattering, while $\Sigma^r_p$ and $i\Gamma_p(E)f_p(E)$ are respectively the retarded and lesser self energies due to lead $p$. Expressions for $\Sigma^r_p$ can be found in the literature\cite{datta}, while $\Gamma_p$ is defined as
\begin{align}\label{gamma_p_defined}
\Gamma_p(E)=i\left(\Sigma_p^r(E)-\Sigma_p^a(E)\right),
\end{align}
where $\Sigma^a_p(E)=\Sigma^r_p(E)^\dag$.

In addition to \refeq{gamma_p_defined}, we also make the definitions
\begin{align}
\Gamma_s(E)=&i\left(\Sigma_s^r(E)-\Sigma_s^a(E)\right),\,\,\text{and}\\
\Gamma(E)=&i\left(\Sigma^r(E)-\Sigma^a(E)\right),
\end{align}
where $\Sigma^a(E)=\Sigma^r(E)^\dag$ and $\Sigma^a_s(E)=\Sigma^r_s(E)^\dag$. It can be shown\cite{datta} that
\begin{align}\label{A_properties}
A(E)=G^r(E)\Gamma(E)G^a(E)=G^a(E)\Gamma(E)G^r(E).
\end{align}

Finally, it can also be shown\cite{datta} that the electric current $I_q$ and heat current $Q_q$ at a lead $q$ will be given respectively as $I_q=-e\int i_q(E)\mathrm{d}E$ and $Q_q=\int(E-\mu_q)i_q(E)\mathrm{d}E$. Here $e$ is the elementary charge, $\mu_q$ the electrochemical potential of lead $q$, while
\begin{align}\label{negfcurrent}
i_q(E)=-\frac{i}{h}\mathrm{Tr}\,\Gamma_q(E)\left(G^<(E)-iA(E)f_q(E)\right).
\end{align}

\subsection{Iterative expansion}\label{it_exp_sec}

The starting point for general Monte Carlo solutions of the Boltzmann equation, is the definition of an iterative solution in the form of a Neumann series. This is possible since the integral form of the Boltzmann equation has the form $f=\mathcal{L}f+f_0$, where $f$ is the Boltzmann distribution function, $\mathcal{L}$ is a linear integral operator, and $f_0$ is an independent source term which arises from the initial conditions\cite{jacoboni}. We can put \refeq{eq_for_gl} into a similar form by introducing the following assumption about $\Sigma^<_s$:

\textbf{Assumption (i)} The lesser scattering self energy can be written as 
\begin{align}\label{lesser_scatt_express}
\Sigma^<_s(E)=\int\mathrm{d}E'\Lambda(E,E',G^<(E')),
\end{align}
where the function $\Lambda(E,E',G)$ is linear in $G$.

This assumption holds at least in the self consistent Born approximation, except for electron-electron scattering. Further justification for the assumption is discussed in appendix \ref{apdix_lin}.

Taking into account \refeq{sigmas_comp_l} and \refeq{lesser_scatt_express}, \refeq{eq_for_gl} becomes
\begin{align}\label{g_less_exp}
G^<(E)&=G^r(E)\left( i\sum_p\Gamma_p(E)f_p(E)+\int\mathrm{d}E'\Lambda(E,E',G^<(E')) \right)G^a(E)\\\nonumber
&=i\sum_pG^r\Gamma_pG^af_p(E)+\int\mathrm{d}E'\Xi(E,E',G^<(E')),
\end{align}
where we have defined
\begin{align}\label{xiint_defined}
\Xi(E,E',G)=G^r(E)\Lambda(E,E',G)G^a(E).
\end{align}
The integral operator $\Xi$ is now defined as
\begin{align}
\Xi G(E)=&\int\mathrm{d}E'\Xi(E,E',G(E')),\label{xi_operator}
\end{align}
and then we write \refeq{g_less_exp} as
\begin{align}\label{g_less_op}
G^<(E)=&\Xi G^<(E)+G^<_0(E),\,\,\text{where}\\\label{g0_less_defined}
G^<_0(E)=&i\sum_pG^r(E)\Gamma_p(E)G^a(E)f_p(E).
\end{align}
The expression for $G_0^<$ can be simplified by making the definitions
\begin{align}\label{ap_defined}
A_p(E)&=G^r(E)\Gamma_p(E)G^a(E),\,\,\text{and}\\
G_p(E)&=A_p(E)f_p(E)=G^r(E)\Gamma_p(E)G^a(E)f_p(E),
\end{align}
so that \refeq{g0_less_defined} can be written
\begin{align}
G^<_0(E)=&i\sum_pG_p(E)=i\sum_pA_p(E)f_p(E).
\end{align}

\refeq{g_less_op} has a form similar to the integral Boltzmann equation. A minor difference is that the source term $G_0^<$ arises from coupling to the leads rather than from initial conditions. A Neumann expansion of $G^<$, similar to what is done with the Boltzmann equation\cite{jacoboni} can now be performed.

However, before we proceed we note that \refeq{g_less_op} depends on the retarded Green's function $G^r$ solving \refeq{eq_for_gr} through \refeq{xiint_defined}. In general, both of the scattering self energies $\Sigma^r_s$ and $\Sigma^<_s$ will depend on both Greens's functions $G^r$ and $G^<$, while $H$ may also depend on $G^<$ through a Poisson potential. Thus, \refeq{eq_for_gr} and \refeq{eq_for_gl} are in fact coupled equations. Since a Neumann expansion based on \refeq{g_less_op} must involve $G^r$, it will generally not be applicable unless both equations have already been solved. This problem will be discussed further in section \ref{lin_limit_section}, and to some extent in section \ref{discuss_hf}. Until then we will simply regard $G^r$ as known.

To find the Neumann expansion of \refeq{g_less_op}, we begin by iteratively defining the functions
\begin{align}\label{itterative_g}
G^<_n(E)&=\Xi G^<_{n-1}(E)+G^<_0(E)=\sum_{m=0}^n\Xi^mG^<_0(E),
\end{align}
where the last step makes use of assumption (i). Assuming that the series converges, the iteration of \refeq{itterative_g} will converge to
\begin{align}\label{gl_neuman_exp}
G^<(E)=\sum_{n=0}^\infty\Xi^nG^<_0(E).
\end{align}
This function solves \refeq{g_less_op}, as can be seen by the evaluation
\begin{align}
\Xi G^<(E)+G^<_0(E)&=\Xi\sum_{n=0}^\infty\Xi^nG^<_0(E)+G^<_0(E)\\\nonumber
&=\sum_{n=0}^\infty\Xi^nG^<_0(E)=G^<(E).
\end{align}
Thus, as indicated by the notation, $G^<(E)$ is in fact the lesser Green's function, and \refeq{gl_neuman_exp} its Neumann expansion.

By now applying \refeq{gl_neuman_exp}, we can then find the energy resolved current of \refeq{negfcurrent} as
\begin{align}\label{negf_it_currexp}
i_q(E)&=-\frac{i}{h}\sum_{n=0}^\infty\mathrm{Tr}\,\Gamma_q(E)\Xi^nG^<_0(E)-\frac{1}{h}\mathrm{Tr}\,\Gamma_q(E)A(E)f_q(E)\\\nonumber
&=\frac{1}{h}\sum_{n=0}^\infty\sum_p\mathrm{Tr}\,\Gamma_q(E)\Xi^nG_p(E)-\frac{1}{h}\mathrm{Tr}\,\Gamma_q(E)A(E)f_q(E).
\end{align}
Since we assume $G^r$ and thus the operator $\Xi$ to be known, \refeq{negf_it_currexp} could in principle be evaluated directly. However, this is computationally demanding for the general case.

\subsection{General Monte Carlo method}\label{general_integral}

The expansion \refeq{negf_it_currexp} is the starting point of the Monte Carlo method we here propose. There are several different ways in which this expression could be evaluated through a Monte Carlo process. Probably, the simplest such method would be to express the sum in terms of matrix elements as
\begin{align}\label{signproblem}
\sum_{n=0}^\infty\sum_p\sum_{i_0j_0\cdots i_nj_n}\int{\mathrm{d}^nE}\,\Gamma^{j_ni_n}_q(E)\Xi^{i_{n-1}j_{n-1}}_{i_nj_n}(E,E_{n-1})\cdots \Xi^{i_0j_0}_{i_1j_1}(E_1,E_0)G^{i_0j_0}_p(E_0),
\end{align}
where $\Gamma^{ij}_q(E)=\langle i|\Gamma_q(E)|j\rangle$, $G^{ij}_p(E)=\langle i|G_p(E)|j\rangle$, and $\Xi^{ij}_{kl}(E,E')=\langle k|\Xi(E,E',|i\rangle\langle j|)|l\rangle$ for some basis $\{|i\rangle\}$. If a set of energies $E_0\cdots E_{n-1}$ and indices $i_0j_0\cdots i_nj_n$ are picked at random from some probability distribution $f$, then the stochastic variable 
\begin{align}\label{scalar_estimator}
x=\frac{\Gamma^{j_ni_n}_q(E)\Xi^{i_{n-1}j_{n-1}}_{i_nj_n}(E,E_{n-1})\cdots \Xi^{i_0j_0}_{i_1j_1}(E_1,E_0)G^{i_0j_0}_p(E_0)}{f}
\end{align}
will have \refeq{signproblem} as its expectation value. Thus, the average of a large number of random values of $x$ could be used as an estimate of the sum in \refeq{negf_it_currexp}. An advantage of this particular approach is that the calculation of each individual Monte Carlo estimate $x$ is numerically fast. It is simply the product of a series of scalar quantities. However, a major challenge is that $x$ is in general a complex quantity. Consequently one can expect instances of $x$ to cancel each other.

Such ``sign problems'' as they are often referred to\cite{PhysRevLett.81.2514}, cause major issues in most quantum Monte Carlo methods. In our case, it makes the use of the estimator of \refeq{scalar_estimator} completely unfeasible, as all cancellations would make the estimate converge very slowly. If possible we should rather try to find an estimator which is always positive in order to remedy the sign problem. From an assumption which will be introduced in the next section, it will follow that the operators in \refeq{negf_it_currexp} are positively definite. Thus, one possibility would be instead of $x$, to use the estimator
\begin{align}
y=\frac{\mathrm{Tr}\,\Gamma_q(E)\Xi^nG^<_0(E)}{p_n},
\end{align}
where the number $n$ is drawn randomly from the probability distribution $p_n$. The estimator $y$ also has an expectation value equal to the sum in \refeq{negf_it_currexp}, but is always positive. The estimator $y$ however, has another problem. The calculation of each estimate is numerically expensive. This is partially because it requires integration over energy, but more importantly because each step of the calculation handles huge and potentially dense matrices $\Xi^{k}G^<_0$, for $k=0, \dotsc, n$.

Thus, since Monte Carlo estimators expressed in terms of scalars and operators both have issues, we shall instead try to follow a middle ground. We seek an estimator expressed in terms of vectors. The easiest way to express \refeq{negf_it_currexp} in terms of vectors, is to diagonalize the operators. For this purpose we introduce an additional assumption about the scattering operators:

\textbf{Assumption (ii) }$\Lambda$ preserves Hermiticity in $G$. That is, for all $E$ and $E'$, and any Hermitian operator $G$, $\Lambda(E,E',G)$ is also Hermitian.

The justification for this assumption is discussed in appendix \ref{apdix_herm}.

Since it can be shown that the operators $G_p(E)=G^r(E)\Gamma_p(E)G^a(E)f_p(E)$ are Hermitian, it follows from assumption (ii) that all of the operators $\Xi(E,E_{n-1},\Xi(\cdots\Xi(E_1,E_0,G_p(E_0))\\\cdots))$ are also Hermitian and diagonalizable. The sum in \refeq{negf_it_currexp} can thus be expressed as
\begin{align}
\sum_{n=0}^\infty\sum_p\int{\mathrm{d}^nE}\,\mathrm{Tr}\,\Gamma_q(E)\sum_{i}\lambda_i^{(n)}|i^{(n)}\rangle\langle i^{(n)}|=\sum_{n=0}^\infty\sum_p\sum_{i}\int{\mathrm{d}^nE}\,\lambda_i^{(n)}\langle i^{(n)}|\Gamma_q(E)|i^{(n)}\rangle,
\end{align}
where $\lambda^{(n)}_i$ and $|i^{(n)}\rangle$ are respectively the eigenvalues and eigenvectors of the operator $\Xi(E,E_{n-1},\Xi(\cdots\Xi(E_1,E_0,G_p(E_0))\cdots))$. It follows from the assumed positive definiteness of these operators that an estimator $\lambda_i^{(n)}\langle i^{(n)}|\Gamma_q(E)|i^{(n)}\rangle/f$ is positive.

While this estimator is expressed in terms of vectors, handling the operators \\$\Xi(E_k,E_{k-1},\Xi(\cdots\Xi(E_1,E_0,G_p(E_0))\cdots))$ for $k=0, \dotsc, n$ is still required during intermediate steps of the calculation. However, this issue is easy to eliminate, since we can start the diagonalization at the innermost operator. Using \refeq{xiint_defined} and the linearity of $\Lambda$, we get
\begin{align}\label{sum_of_omegas_3}
&\Xi(E,E_{n-1},\Xi(\cdots\Xi(E_1,E_0,G_p(E_0))\cdots))\\\nonumber
=&\Xi(E,E_{n-1},\Xi(\cdots\Xi(E_1,E_0,G^r(E_0)\Gamma_p(E_0)G^a(E_0))\cdots))f_p(E_0)\\\nonumber
=&\sum_{i_0}\Xi(E,E_{n-1},\Xi(\cdots G^r(E_1)\Lambda(E_1,E_0,\lambda_{i_0}G^r(E_0)|i_0\rangle\langle i_0|G^a(E_0))G^a(E_1)\cdots))f_p(E_0)\\\nonumber
=&\sum_{i_0, i_1}\Xi(E,E_{n-1},\Xi(\cdots \lambda_{i_1}G^r(E_1)|i_1\rangle\langle i_1|G^a(E_1)\cdots))\lambda_{i_0}f_p(E_0)\\\nonumber
=&\cdots\\\nonumber
=&\sum_{i_0, i_1, \dotsc, i_{n-1}}G^r(E)\Lambda\left(E,E_{n-1},G^r(E_{n-1})|i_{n-1}\rangle\langle i_{n-1}|G^a(E_{n-1})\right)G^a(E)\lambda_{i_0}\lambda_{i_1}\cdots \lambda_{i_{n-1}}f_p(E_0)\\\nonumber
=&\sum_{i_0, i_1, \dotsc, i_{n}}G^r(E)|i_n\rangle\langle i_n|G^a(E)\lambda_{i_0}\lambda_{i_1}\cdots \lambda_{i_{n}}f_p(E_0),\\\nonumber
\end{align}
where $\lambda_{i_k}$ and $|i_k\rangle$ are now respectively the eigenvalues and eigenvectors of the operators $\Lambda^{(k)}$, defined recursively as
\begin{align}\label{lambda_k_defined}
\Lambda^{(k)}=\Lambda\left(E_k,E_{k-1},G^r(E_{k-1})|i_{k-1}\rangle\langle i_{k-1}|G^a(E_{{k-1}})\right)
\end{align}
for $k \geq 1$, and as $\Lambda^{(0)}=\Gamma_p(E_0)$.

Inserting \refeq{sum_of_omegas_3} into \refeq{negf_it_currexp}, we get
\begin{align}\label{i_from_lambda_expansion}
i_q(E)=&\frac{1}{h}\sum_p\sum_{n=0}^\infty\sum_{i_0, i_1, \dotsc, i_{n}}\int\mathrm{d}^nE\,\langle i_n|G^a(E)\Gamma_q(E)G^r(E)|i_n\rangle\lambda_{i_0}\lambda_{i_1}\cdots \lambda_{i_{n}}f_p(E_0)\\\nonumber
&-\frac{1}{h}\mathrm{Tr}\,\Gamma_q(E)A(E)f_q(E).
\end{align}
Now consider a probability distribution $g(p, n, \boldsymbol{\phi}, \boldsymbol{E}, r)$ from which we draw an initial lead $p$, a final lead $r$, a number $n$, a vector of indices $\boldsymbol{\phi}=[i_0, i_1, \dotsc, i_n]$, and a vector of energies $\boldsymbol{E}=[E_0, \dotsc, E_{n-1}, E_n]$. We define a stochastic variable
\begin{align}\label{mc_X_def}
X_q&=\frac{f_p(E_0)\lambda_{i_0}\lambda_{i_1}\cdots \lambda_{i_{n}}\langle i_n|G^a(E_n)\Gamma_r(E_n)G^r(E_n)|i_n\rangle}{g(p, n, \boldsymbol{\phi}, \boldsymbol{E}, r)}\delta_{rq},
\end{align}
the expectation value of which will be
\begin{align}\label{mc_exp_sum}
\langle X_q\rangle=&\sum_{prn\boldsymbol{\phi}}\int\mathrm{d}^{n+1}E\,X_q(p, n, \boldsymbol{\phi}, \boldsymbol{E}, r)g(p, n, \boldsymbol{\phi}, \boldsymbol{E}, r)\\\nonumber
=&\sum_{pn\boldsymbol{\phi}}\int\mathrm{d}^{n+1}E\,f_p(E_0)\lambda_{i_0}\lambda_{i_1}\cdots \lambda_{i_{n}}\langle i_n|G^a(E_n)\Gamma_q(E_n)G^r(E_n)|i_n\rangle.
\end{align}

\refeq{mc_exp_sum} is seen to be identical to the sum of \refeq{i_from_lambda_expansion}, except that the integral is also over the final energy $E_n$. Thus, we can estimate the current at lead $q$ as
\begin{align}\label{negfmc_i_exp}
I_q=-e\int\mathrm{d}E_n\,i_q(E_n)=-\frac{e}{h}\left(\bar{X_q}-\int\mathrm{d}E\,\mathrm{Tr}\,A(E)\Gamma_qf_q(E)\right),
\end{align}
where $\bar{X_q}$ denotes the average of $X_q$ over the Monte Carlo ensemble. The heat current $Q_q$ can be evaluated by an analogous expression.

The assumption of positive definiteness leads to a positive estimator $X_q$. We thus avoid the sign problem. In addition, the calculation of $X_q$ has the desired property of mainly involving vector operations. The only exception to this is that one must diagonalize the operators $\Lambda^{(k)}$. However, given reasonable approximations one can assume these operators to be sparse. This approach is then still significantly more efficient compared to dealing with the operators $\Xi^{k}G^<_0$, which are likely to be dense for sufficiently high $k$. In fact, by utilizing an additional approximation in section \ref{buttiker_section}, we can avoid diagonalization all together. However, let us first define an algorithm which determines the probability distribution $g$.

\subsection{Scattering Monte Carlo method}\label{intuitive_integral}

In the literature\cite{lundstrom,jacoboni} one will find described a particularly intuitive BMC method, where electrons are scattered from state to state with probabilities proportional to the physical scattering rates. While improvements in terms of convergence can be made using so called weighted Monte Carlo methods\cite{jacoboni}, the benefit of the unweighted intuitive approach is that the estimators have particularly simple expressions. Generally, one should expect complicated estimators to have smaller variance than simpler ones only in carefully constructed cases. So as a starting point, we in this work seek to mimic the intuitive scattering Monte Carlo method, in the hope that this will result in a moderate variance.

We start by interpreting the function $\Lambda$ as somewhat analogous to the scattering rate $\Gamma$ in the Boltzmann equation, and seek to express the probability of scattering to new states in terms of $\Lambda$. The involved quantities need to be positive and we need to introduce an assumption about positive definiteness. While we have briefly mentioned such an assumption earlier, in relation to discussions of the sign problem, we now introduce it formally.

\textbf{Assumption (iii) }$\Lambda$ preserves positive definiteness in $G$. That is, for all $E$ and $E'$, and any Hermitian positively definite operator $G$, $\Lambda(E,E',G)$ is also positively definite.

The justification for this assumption is discussed in appendix \ref{apdix_pos}. We make use of assumption (iii), and define the following Monte Carlo procedure:

\textit{Step 1.} Draw a starting lead $p$, all leads having equal probability $1/N$.

\textit{Step 2.} Draw a starting energy $E_0$ from some probability distribution $h(E)$.

\textit{Step 3.} Diagonalize $\Gamma_p(E_0)$ to find its eigenvectors $|i\rangle$ and eigenvalues $\lambda_i$.

\textit{Step 4.} We construct a probability distribution
\begin{align}\label{lead_out_probability}
p_i=\frac{\lambda_i\langle i|A(E_0)|i\rangle}{\mathrm{Tr}\,\Gamma_p(E_0) A(E_0)},
\end{align}
From the expressions for $\Gamma_p$ found in the literature\cite{datta}, we see that it is positive definite. Furthermore, the spectral density $A$ must also be positively definite due to its relation to the density of states $D_i(E)=\langle i|A(E)|i\rangle$. Thus, $p_i$ is positive and sums to one. Draw an index $i_0$ from this probability distribution, and set $n=0$.

{Heuristic justification:} The probability that an electron coming from lead $p$ will enter the system in state $|i\rangle$, is proportional to the projected density of states in $|i\rangle$, and to the strength of interaction between the state $|i\rangle$ and lead $p$. The density of states in $|i\rangle$ is given by $D_i(E)=\langle i|A(E)|i\rangle$, while the mentioned interaction strength is represented by the eigenvalues $\lambda_i$. The requirement of summation to one gives \refeq{lead_out_probability}.

\textit{Step 5.} Find the state vector $|\psi_n\rangle=G^r(E_n)|i_n\rangle$. It can be found either by explicitly calculating the retarded Green's function $G^r$, or alternatively by solving the modified Schrodinger equation
\begin{align}\label{modified_schrodinger}
\left(E_n-H-\Sigma^r(E_n)\right)|\psi_n\rangle=|i_n\rangle.
\end{align}
{Heuristic justification:} This step represents the free propagation of electrons between scattering events.

\textit{Step 6.} Calculate the quantities $\mathrm{Tr}\,\Gamma_r|\psi_n\rangle\langle \psi_n|=\langle \psi_n|\Gamma_r|\psi_n\rangle$ for all leads $r$, and also
\begin{align}\label{quantity_F}
S_n=\int\mathrm{d}E\,\mathrm{Tr}\,A(E)\Lambda\left(E,E_n,|\psi_n\rangle\langle \psi_n|\right).
\end{align}

\textit{Step 7.} From \refeq{lambda_k_defined}, we recognize
\begin{align}\label{diagonal_lambda_1}
\Lambda\left(E,E_n,|\psi_n\rangle\langle \psi_n|\right)=\Lambda\left(E,E_n,G^r(E_n)|i_n\rangle\langle i_n|G^a(E_n)\right)=\Lambda^{(n+1)}=\sum_i\lambda_{i}|i\rangle\langle i|.
\end{align}
From assumption (iii), the eigenvalues $\lambda_i$ are positive, so that
\begin{align}\label{ALambda_is_positive}
\mathrm{Tr}\,A(E)\Lambda\left(E,E_n,|\psi_n\rangle\langle \psi_n|\right)=\sum_i\lambda_{i}\langle i|A(E)|i\rangle\geq 0.
\end{align}
Accordingly, $S_n$ is also positive, so we can construct a probability distribution
\begin{align}\label{lead_out_probability_buttiker}
p_r=\frac{\langle \psi_n|\Gamma_r(E_n)|\psi_n\rangle}{\sum_b\langle \psi_n|\Gamma_b(E_n)|\psi_n\rangle+S_n}.
\end{align}
Attempt to draw a lead $r$ from this distribution, which is positive, but does not sum to one. Since the distribution does not sum to one, a successful draw is not guaranteed. If the draw is successful and a lead $r$ is drawn, stop the process and choose lead $r$ as the end lead. Otherwise continue the process.

{Heuristic justification:} After free propagation, the electron will either scatter to a new state, or exit the system through one of the leads. Following the reasoning behind step 4, the probability of exiting through a lead $r$ should be proportional to $\mathrm{Tr}\,\Gamma_r|\psi_n\rangle\langle \psi_n|=\langle \psi_n|\Gamma_r|\psi_n\rangle$. By interpreting $A(E)$ as the density of states, and the function $\Lambda$ as describing the scattering interaction strength, one could argue that the scattering rate to energy $E$ should be given by $\mathrm{Tr}\,A(E)\Lambda\left(E,E_n,|\psi_n\rangle\langle \psi_n|\right)$. Thus, $S_n$ given by \refeq{quantity_F} represents the total scattering rate. Since the probability of scattering is proportional to $S_n$, while the probabilities of exiting through a lead $r$ is proportional to $\langle \psi_n|\Gamma_r|\psi_n\rangle$, the requirement of total summation to one gives \refeq{lead_out_probability_buttiker}.

\textit{Step 8.} Draw an energy $E_{n+1}$ from the probability distribution
\begin{align}\label{mc_directe_pofe}
p(E)=\frac{\mathrm{Tr}\,A(E)\Lambda\left(E,E_n,|\psi_n\rangle\langle \psi_n|\right)}{S_n}.
\end{align}
which integrates to 1, and is positive from \refeq{ALambda_is_positive}.

{Heuristic justification:} $\mathrm{Tr}\,A(E)\Lambda\left(E,E_n,|\psi_n\rangle\langle \psi_n|\right)$ represents the scattering rate to energy $E$, while $S_n$ represents the total scattering rate.

\textit{Step 9.} Perform the diagonalization of \refeq{diagonal_lambda_1} to find the eigenvectors $|i\rangle$ and eigenvalues $\lambda_i$ of $\Lambda^{(n+1)}=\Lambda\left(E_{n+1},E_n,|\psi_n\rangle\langle \psi_n|\right)$. By assumption (iii) these are all positive.

\textit{Step 10.} Draw an index $i_{n+1}$ from the distribution
\begin{align}\label{mc_p_newi}
p_i=\frac{\lambda_i\langle i|A(E_{n+1})|i\rangle}{\mathrm{Tr}\,A(E_{n+1})\Lambda^{(n+1)}}=\frac{\lambda_i\langle i|A(E_{n+1})|i\rangle}{\mathrm{Tr}\,A(E_{n+1})\Lambda\left(E_{n+1},E_n,|\psi_n\rangle\langle \psi_n|\right)}.
\end{align}

{Heuristic justification:} Following again the reasoning of step 4, the probability for an electron to scatter from state $|\psi_n\rangle$ to $|i\rangle$, is proportional to the projected density of states in $|i\rangle$, and to the strength of interaction between $|i\rangle$ and $|\psi_n\rangle$. The density of states in $|i\rangle$ is given by $D_i(E)=\langle i|A(E)|i\rangle$, while the interaction strength is represented by the eigenvalues $\lambda_i$ from step 9. Requiring summation to one we get \refeq{mc_p_newi}.

\textit{Step 11.} Increment $n$ by 1, and return to step 5. Continue the procedure from there.

The steps 1 to 11 define the Monte Carlo procedure. The procedure is truncated only when an end lead $r$ is drawn in step 7. At that point we will have selected start and end leads $p$ and $r$, and a set of $n+1$ energies and $n+1$ indices. Thus, the procedure defines a probability distribution $g(p, n, \boldsymbol{\phi}, \boldsymbol{E}, r)$.

We can evaluate this probability by taking the product of the probabilities associated with each step of the procedure. Clearly, steps 1 to 4 always give rise to a combined factor of
\begin{align}\label{1_to_4_prob}
\frac{h(E_0)}{N}\frac{\lambda_{i_0}\langle i_0|A(E_0)|i_0\rangle}{\mathrm{Tr}\,\Gamma_p(E_0) A(E_0)}.
\end{align}
The contributions from the remaining steps will depend on whether the electron scatters or exits at a lead in step $7$. The probability of scattering is the same as the probability of not exiting at a lead, so that every scattering event will contribute a factor
\begin{align}
1-\sum_rp_r=\frac{S_n}{\sum_b\langle \psi_n|\Gamma_b(E_n)|\psi_n\rangle+S_n},
\end{align}
to the total probability, $p_r$ being given by \refeq{lead_out_probability_buttiker}. In addition, each scattering event will also contribute a factor $p(E)$ from \refeq{mc_directe_pofe}, and a factor $p_i$ from \refeq{mc_p_newi}. Thus, assuming that the electron is scattered $n$ times before exiting, the total contribution to $g$ from scattering becomes 
\begin{align}\label{total_scatter_prob}
\prod_{k=0}^{n-1}\,\,&\frac{S_k}{\sum_b\langle \psi_k|\Gamma_b(E_k)|\psi_k\rangle+S_k}\times\frac{\mathrm{Tr}\,A(E_{k+1})\Lambda\left(E_{k+1},E_k,|\psi_k\rangle\langle \psi_k|\right)}{S_k}\\\nonumber
\times&\frac{\lambda_{i_{k+1}}\langle i_{k+1}|A(E_{k+1})|i_{k+1}\rangle}{\mathrm{Tr}\,A(E_{k+1})\Lambda\left(E_{k+1},E_k,|\psi_k\rangle\langle \psi_k|\right)}=\prod_{k=0}^{n-1}\,\frac{\lambda_{i_{k+1}}\langle i_{k+1}|A(E_{k+1})|i_{k+1}\rangle}{\sum_b\langle \psi_k|\Gamma_b(E_k)|\psi_k\rangle+S_k}
\end{align}

The probability of finally exiting at lead $r$ after the $n$ scattering events, is given by \refeq{lead_out_probability_buttiker}. Thus, taking the product of \refeq{1_to_4_prob}, \refeq{total_scatter_prob} and \refeq{lead_out_probability_buttiker}, we obtain
\begin{align}\label{intuitive_g_probability}
&g(p, n, \boldsymbol{\phi}, \boldsymbol{E}, r)\\\nonumber
=&\frac{h(E_0)}{N}\frac{\lambda_{i_0}\langle i_0|A(E_0)|i_0\rangle}{\mathrm{Tr}\,\Gamma_p(E_0) A(E_0)}\times\prod_{k=0}^{n-1}\,\frac{\lambda_{i_{k+1}}\langle i_{k+1}|A(E_{k+1})|i_{k+1}\rangle}{\sum_b\langle \psi_k|\Gamma_b(E_k)|\psi_k\rangle+S_k}\times\frac{\langle \psi_n|\Gamma_r(E_n)|\psi_n\rangle}{\sum_b\langle \psi_n|\Gamma_b(E_n)|\psi_n\rangle+S_n}\\\nonumber
=&\frac{h(E_0)}{N}\frac{\langle \psi_n|\Gamma_r(E_n)|\psi_n\rangle}{\mathrm{Tr}\,\Gamma_p(E_0)A(E_0)}\prod_{k=0}^{n}\frac{\lambda_{i_k}\langle i_k|A(E_k)|i_k\rangle}{\sum_b\langle \psi_k|\Gamma_b(E_k)|\psi_k\rangle+S_k}.
\end{align}
Having found an explicit expression for the probability distribution $g$, we can find an expression for the estimator $X_q$ using \refeq{mc_X_def}. Inserting \refeq{intuitive_g_probability} in \refeq{mc_X_def}, we find that the eigenvalues and the quantity $\langle \psi_n|\Gamma_r(E_n)|\psi_n\rangle=\langle i_n|G^a(E_n)\Gamma_r(E_n)G^r(E_n)|i_n\rangle$ cancel, and what remains is
\begin{align}\label{X_def_intmc}
X_q=N\cdot\frac{f_p(E_0)}{h(E_0)}\cdot\mathrm{Tr}\,\Gamma_p(E_0)A(E_0)\cdot\prod_{k=0}^n\frac{\sum_b\langle \psi_k|\Gamma_b(E_k)|\psi_k\rangle+S_k}{\langle i_k|A(E_k)|i_k\rangle}\delta_{rq}.
\end{align}
While these cancellations simplifies the expression, what remains is still fairly involved. However, as we will see, \refeq{X_def_intmc} can be simplified even further by making an additional assumption about $\Lambda(E',E,G)$:

\textbf{Assumption (iv)} For any Hermitian and positively definite operator $G$,
\begin{align}\label{lesser_scatt_charge}
\mathrm{Tr}\,\Gamma_s(E)G=\int\mathrm{d}E'\,\mathrm{Tr}\,A(E')\Lambda(E',E,G).
\end{align}

Assumption (iv) can be interpreted as an assumption of the scattering mechanism being charge conserving. This is more thoroughly discussed in appendix \ref{apdix_cc}.

Employing assumption (iv), we first note that in step 6 above, we can calculate the quantity $S_n$ of \refeq{quantity_F} simply as
\begin{align}\label{charge_cons_F}
S_n=\mathrm{Tr}\,\Gamma_s(E_n)|\psi_n\rangle\langle\psi_n|=\langle\psi_n|\Gamma_s(E_n)|\psi_n\rangle,
\end{align}
so that we no longer need to perform the integral explicitly. Next, making use of \refeq{A_properties} and \refeq{sigmas_comp_r}, we see that
\begin{align}
\sum_b\langle \psi_n|\Gamma_b(E_n)|\psi_n\rangle+S_n&=\sum_b\langle \psi_n|\Gamma_b(E_n)|\psi_n\rangle+\langle\psi_n|\Gamma_s(E_n)|\psi_n\rangle=\langle\psi_n|\Gamma(E_n)|\psi_n\rangle\\\nonumber
&=\langle i_n|G^a(E_n)\Gamma(E_n)G^r(E_n)|i_n\rangle=\langle i_n|A(E_n)|i_n\rangle.
\end{align}
Inserting this into \refeq{X_def_intmc}, we see that the remaining terms in the product also cancel, so that $X_q$ becomes simply
\begin{align}\label{X_def_intmc_cc}
X_q=\frac{N\cdot f(E_0)\cdot\mathrm{Tr}\,\Gamma_p(E_0)A(E_0)}{h(E_0)}\delta_{rq},
\end{align}
which has the simple form we were looking for.

\subsection{Linear elastic limit}\label{lin_limit_section}

In the preceding discussions we assumed that either the retarded Green's function $G^r$, or at least its inverse is known. In the previous section for instance, an expression for $G^r$ or its inverse was required in step 5. But in reality $G^r$ is rarely known beforehand, since the scattering self energy $\Sigma^r_s$ usually has a functional dependence on $G^<$. In fact, even the Hamiltonian $H$ may depend on $G^<$, for instance through a Poisson potential. Thus, \refeq{eq_for_gr} and \refeq{eq_for_gl} must generally be solved self consistently with each other. The method presented above will thus not be applicable. In section \ref{discuss_hf} we briefly outline a possible approach for how to deal with this in the general case. In this section we restrict the discussion to linear transport.

In the linear limit, the existence of Kubo expressions tells us that the currents should be expressible in terms of equilibrium quantities. If we could replace $G^r$ above with its equilibrium value, then \refeq{eq_for_gr} and \refeq{eq_for_gl} would no longer be coupled, and the described method would thus be applicable. While it is difficult to show that this substitution is possible in the general case, it is rather straightforward if we also assume elastic scattering. In that case, we have
\begin{align}\label{elastic_def}
\Lambda(E',E,G)=\Lambda(E,G)\delta(E-E'),
\end{align}
so that the operator $\Xi$ is given simply by
\begin{align}
\Xi G(E)&=G^r(E)\Lambda(E,G(E))G^a(E)\label{elastic_xi_op}
\end{align}
In addition, we need one final assumption about the scattering mechanism, namely

\textbf{Assumption (v)} When the scattering mechanism is elastic, the function $\Lambda(E,G)$ satisfies
\begin{align}
\mathrm{Tr}\,F\Lambda(E,G)=\mathrm{Tr}\,G\Lambda(E,F),
\end{align}
where $F$ and $G$ are any two Hermitian and positively definite operators.

Intuitively, this assumption can be interpreted as a statement of detailed balance. Its justification is discussed in appendix \ref{apdix_detb}.

It follows from assumptions (v) and (iv) that the quantity $\mathrm{Tr}\,A(E)\Gamma_q(E)$ can be expanded as

\begin{align}\label{detail_b_aexp}
\mathrm{Tr}\,A(E)\Gamma_q(E)=\sum_{n=0}^\infty\sum_p\mathrm{Tr}\,\Gamma_q(E)\Xi^nA_p(E).
\end{align}
The proof of this is given in appendix \ref{apdix_detailed}. Inserting \refeq{detail_b_aexp} into \refeq{negf_it_currexp} yields
\begin{align}\label{iqof_t_el_db}
i_q(E)&=\frac{1}{h}\sum_p\sum_{n=0}^\infty\mathrm{Tr}\,\Gamma_q(E)\Xi^nA_p(E)\left(f_p(E)-f_q(E)\right)\\\nonumber
&\equiv\frac{1}{h}\sum_p\mathcal{T}_{qp}(E)\left(f_p(E)-f_q(E)\right),
\end{align}
where $\mathcal{T}_{qp}(E)=\sum_{n=0}^\infty\mathrm{Tr}\,\Gamma_q(E)\Xi^nA_p(E)$.
Assume now that the the system's offset from equilibrium is described by some parameter $x$, which could for instance be a voltage or a temperature difference between two of the leads. Then all functions in \refeq{iqof_t_el_db} are functions also of $x$. Expanding to first order in $x$ yields
\begin{align}\label{iq_lin_elast_1}
i_q(E)=&\sum_p\frac{\partial\mathcal{T}_{qp}}{\partial x}(E)\left(f_p(E)-f_q(E)\right)x+\sum_p\mathcal{T}_{qp}(E)\left(\frac{\partial f_p}{\partial x}(E)-\frac{\partial f_q}{\partial x}(E)\right)x,
\end{align}
where all functions on the right are evaluated at equilibrium. But in equilibrium $f_p(E)=f_q(E)$, so that in fact
\begin{align}\label{iq_lin_elast_2}
i_q(E)&=\sum_p\mathcal{T}^{Eq}_{qp}(E)\Delta f_{pq}(E),\,\,\text{where}\\
\Delta f_{pq}(E)&=\left(\frac{\partial f_p}{\partial x}(E)-\frac{\partial f_q}{\partial x}(E)\right)x.
\end{align}
Thus, all transport coefficients can be evaluated from the transmission functions $\mathcal{T}^{Eq}_{qp}(E)$ at equilibrium. The conductance $G_{qp}$ for instance, are given by the expression
\begin{align}\label{Gqp_expression}
G_{qp}=\frac{e^2}{h}\int\mathrm{d}E\,\mathcal{T}^{Eq}_{qp}(E)\mathrm{Th}(E),
\end{align}
where $\mathrm{Th}(E)=-\partial f/\partial E$. This means that $G_{qp}$ can be estimated as
\begin{align}\label{gpq_formula}
G_{qp}=\frac{e^2}{h}\bar{X}_{pq},
\end{align}
where the stochastic variables $X_{pq}$ are found using the Monte Carlo procedure described above with a few modifications; the initial lead $p$ is kept fixed, $f(E)$ is replaced by $\mathrm{Th}(E)$, and $G^r$ and the function $\Lambda$ are replaced with their respective equilibrium values. If we also choose the probability distribution as $h(E)=\mathrm{Th}(E)$, then \refeq{X_def_intmc_cc} becomes simply
\begin{align}\label{elast_lin_xpq}
X_{pq}=\mathrm{Tr}\,\Gamma_p(E_0)A(E_0)\delta_{rq}.
\end{align}

\subsection{B\"uttiker's Approximation}\label{buttiker_section}

In the general case, the bottleneck of the procedure described in section \ref{intuitive_integral} will be step 9, which requires diagonalization of the operators $\Lambda^{(n+1)}$. In order for the procedure to be applicable in practice, one must probably always find some approximation which improves the calculation efficiency of this step. The simplest of such is to assume that $\Lambda^{(n+1)}$ is already diagonal in the adopted basis. In fact, the procedure can be simplified even further by making some additional assumptions. First, we assume the existence of a single basis set $\{|i\rangle\}$ which diagonalizes both $\Sigma^r(E)$ and $\Lambda(E,E',G)$ for all $E$, $E'$ and all Hermitian and positively definite operators $G$. Then, that the eigenvalue $\lambda_i$ of $\Lambda(E,E',G)$ corresponding to the basis vector $|i\rangle$, is a function only of the matrix element $\langle i|G|i\rangle$. Or more compactly
\begin{align}\label{buttiker_scattering_gamma}
\Sigma^r(E)&=\sum_i\sigma_i(E)|i\rangle\langle i|,\\\label{buttiker_scattering_sigma}
\Lambda(E,E',G)&=\sum_i\lambda_i(E,E',\langle i|G|i\rangle)|i\rangle\langle i|.
\end{align}

We refer to \refeq{buttiker_scattering_gamma} and \refeq {buttiker_scattering_sigma} as B\"uttiker's approximation. The reason for this is that these are the self energies we obtain if we model electron scattering as a set of floating voltage probes attached to the states $|i\rangle$. The idea of modeling scattering with such probes was introduced by B\"uttiker\cite{buttiker}, and the probes are often referred to as B\"uttiker probes\cite{bulusu2008review}.

If the scattering model defined by \refeq{buttiker_scattering_gamma} and \refeq{buttiker_scattering_sigma} is to satisfy assumption (i), then the functions $\lambda_i$ must be linear in the third argument, so that $\lambda_i(E,E',\langle i|G|i\rangle)=\tilde{\lambda}_i(E,E')\langle i|G|i\rangle$. If assumptions (ii) and (iii) are also to be satisfied, the functions $\tilde{\lambda}_i$ must be real and positive: $\tilde{\lambda}_i(E,E')\geq 0$. Finally, if the scattering model is to satisfy condition (iv), then we must have
\begin{align}\label{condition4_buttiker}
i(\sigma_i-\sigma_i^\star)\equiv\gamma_i(E)=\int\mathrm{d}E\,\tilde{\lambda}_i(E',E)\langle i|A(E')|i\rangle.
\end{align}

If we in addition to B\"uttiker's approximation also assume elastic scattering, then $\tilde{\lambda}_i(E,E')=\tilde{\lambda}(E)\delta(E-E')$, and \refeq{condition4_buttiker} becomes
\begin{align}
\gamma_i(E)=\tilde{\lambda}_i(E)\langle i|A(E')|i\rangle.
\end{align}
In that case, we also see that
\begin{align}
\mathrm{Tr}\,F\Lambda(E,G)=\sum_i\langle i|F|i\rangle\tilde{\lambda}_i(E)\langle i|G|i\rangle=\mathrm{Tr}\,G\Lambda(E,F),
\end{align}
so that assumption (v) is satisfied automatically.

We are now at a point where the procedure of section \ref{intuitive_integral} can be greatly simplified. First of all, since $\Lambda\left(E,E_n,|\psi_n\rangle\langle \psi_n|\right)$ is already diagonal in the basis $\{|i\rangle\}$, step 9 can be omitted. Secondly, the total probability associated with steps 8 and 10 will be
\begin{align}\label{total_p_iE}
p_{i}(E)=&\frac{\sum_i\tilde{\lambda}_i(E,E_n)|\langle\psi_n|i\rangle|^2\langle i|A(E)|i\rangle}{S_n}\times\frac{\tilde{\lambda}_i(E,E_n)|\langle\psi_n|i\rangle|^2\langle i|A(E)|i\rangle}{\sum_i\tilde{\lambda}_i(E,E_n)|\langle\psi_n|i\rangle|^2\langle i|A(E)|i\rangle}\\\nonumber
=&\frac{\tilde{\lambda}_i(E,E_n)|\langle\psi_n|i\rangle|^2\langle i|A(E)|i\rangle}{S_n}.
\end{align}
Accordingly, we can reverse the order of these two steps by first picking an index $i$ with probability
\begin{align}\label{draw_i_first}
p_{i}=\frac{\gamma_i|\langle\psi_n|i\rangle|^2}{\sum_i\gamma_i|\langle\psi_n|i\rangle|^2}=\frac{\gamma_i|\langle\psi_i|i\rangle|^2}{S_n}.
\end{align}
and then an energy $E$ with probability
\begin{align}\label{draw_E_last}
p(E)=\frac{\tilde{\lambda}_i(E,E_n)\langle i|A(E)|i\rangle}{\gamma_i}=\frac{\tilde{\lambda}_i(E,E_n)\langle i|A(E)|i\rangle}{\int\mathrm{d}E\,\tilde{\lambda}_i(E,E_n)\langle i|A(E)|i\rangle},
\end{align}
where we have used \refeq{condition4_buttiker} and \refeq{charge_cons_F}. The product of the two probabilities \refeq{draw_i_first} and \refeq{draw_E_last} results in the same total probability $p_i(E)$ as \refeq{total_p_iE}.

Finally, the step of picking an exit lead $r$ (step 7) or a new internal index $i$, can be combined into a single step where leads are still chosen according to \refeq{lead_out_probability_buttiker}, but internal indices are picked with probability
\begin{align}\label{draw_i_simult}
p_i=\frac{\gamma_i|\langle\psi_n|i\rangle|^2}{\sum_r\langle \psi_n|\Gamma_r|\psi_n\rangle+\sum_j\gamma_j|\langle\psi_n|j\rangle|^2}.
\end{align}
The probability of picking a particular index $i$, \textit{given} that a lead is not drawn, is then given precisely by \refeq{draw_i_first}.

Thus, steps 6 to 10 of section \ref{intuitive_integral}, can be replaced by the steps

\textit{Step 6'} Draw an internal index $i_{n+1}$ with probabilities given by \refeq{draw_i_simult}, or a lead $r$ with probabilities
\begin{align}
p_r=\frac{\langle \psi_n|\Gamma_r|\psi_n\rangle}{\sum_r\langle \psi_n|\Gamma_r|\psi_n\rangle+\sum_j\gamma_j|\langle\psi_n|j\rangle|^2}.
\end{align}
If a lead is drawn, stop the process. If an internal index is drawn, continue.

\textit{Step 7'} Draw an energy $E_{n+1}$ from the probability distribution of \refeq{draw_E_last}.

After step 7', the process is continued from step 11.

\section{Model and implementation}\label{implem}

\subsection{Model}

Our tests of the described method have been performed within B\"uttiker's approximation, using the specialized method described in section \ref{buttiker_section}. Application to more general scattering mechanisms, using the method of section \ref{intuitive_integral}, is postponed for future studies. In addition, the performed tests assume elastic scattering. Step 7' of section \ref{buttiker_section} is thus omitted, while the linear conductances $G_{qp}$ are calculated using \refeq{gpq_formula}. The energies $E_0$ of the Monte Carlo samples are chosen from the probability distribution $h(E)=\mathrm{Th}(E)$, so that the sampled variable $X_{pq}$ is given by \refeq{elast_lin_xpq}.

In addition to the scattering operators $\Sigma^r$ and $\Sigma^<$, we need a model Hamiltonian $H$. For this purpose, we make use of an atomistic tight binding model of the material CdTe. The model is based on previous literature\cite{PhysRevB.37.8215} which parameterizes nearest neighbor interaction in sp3-hybridized fcc-structures in terms of seven parameters. We find these parameters by fitting against DFT calculations. The tight binding model and fitting procedure is described in more detail in our previous work\cite{paper1}. The geometries are described in section \ref{results}.

Since step 7' is omitted, the only information we require of the scattering model, is the basis set $\{|i\rangle\}$ in which the scattering operators are diagonal, and the functions $\sigma_i(E)$. The basis set $\{|i\rangle\}$ is chosen to be the s and p orbitals employed in the tight binding model, while we choose $\sigma_i(E)$ to be purely imaginary, so that 
\begin{align}
\sigma_i(E)=-\frac{i}{2}\gamma_i(E)=-\frac{i\hbar}{2\tau}.
\end{align}
In general the lifetime $\tau$ could be a function of both $E$ and $i$, but in our tests it is set to a constant.

All tests are performed with only two leads $p=1$ and $p=2$. To simplify the procedure we write the lead self energies as
\begin{align}
\Sigma^r_p(E)=-\frac{i}{2}\Gamma_lP_p,
\end{align}
where $P_p$ is a projection operator projecting onto orbitals on the boundary with lead $p$, and $\Gamma_l$ is a constant describing the strength of coupling between the system and leads. This is an unrealistic model of $\Sigma^r_p(E)$, but the physics relevant to the tests remains correct.

\subsection{Monte Carlo implementation}

Our prototype implementation follows the approach described in section \ref{buttiker_section}. It is written in Python, and makes use of the RGF algorithm\cite{lake1997single} to calculate $G^r$ by solving \refeq{eq_for_gr}. Since the basis set $\{|i\rangle\}$ is the tight binding basis set, we only need a single column of $G^r$ for each execution of step 5. Thus, the diagonal of $G^r$ is calculated in the beginning of each simulation, while the required columns of $G^r$ are calculated as needed, upon execution of step 5.

The instantiation of stochastic variables is done by transforming to a homogeneously distributed variables\cite{press2007numerical}, which are drawn using the Numpy rand function. In particular, the execution of step 6' is done as follows:

An array $a_j$ is defined, storing the values $\gamma_i|\langle\psi_n|i\rangle|^2$, and $\langle \psi_n|\Gamma_r|\psi_n\rangle$. The cumulative sum $b_j$ of the array $a_j$ is calculated. $b_j$ is then divided by its last element, which normalizes it to a cumulative probability distribution. A random number $x$ is picked homogeneously in the range $[0,1]$, and the Numpy argmax function is used to find the first value in $b_j$ which is larger than $x$. If the index $j$ corresponds to an element of $a_j$ given by $\gamma_i|\langle\psi_n|i\rangle|^2$, then the internal index $i_{n+1}=i$ is picked. If the index $j$ corresponds to an element of $a_j$ given by $\langle \psi_n|\Gamma_r|\psi_n\rangle$, then the lead $r$ is picked as an exit point.

\subsection{Direct and iterative methods}\label{direct_methods}

For comparison reasons we also implemented a scheme where \refeq{eq_for_gl} is solved directly. Within the elastic B\"uttiker approximation, \refeq{eq_for_gl} can be written
\begin{align}\label{buteq}
i_j(E)=\sum_p\langle j|G^r\Sigma^<_pG^a|j\rangle+\sum_i\gamma_i\gamma_j\left|G^r_{ji}(E)\right|^2i_i(E),
\end{align}
where $i_j(E)=\lambda_i(E)\langle i|A(E)|i\rangle=\tilde{\lambda}_i(E)A_{ii}(E)G^<_{ii}(E)$. \refeq{buteq} can be solved directly or iteratively, and we test both of these against the Monte Carlo method. The direct method employs the Numpy solve function, while the iterative method employs the Scipy implementation of the gmres method.

To evaluate the transport coefficients $G_{pq}$, we must integrate the quantity $i_p(E)$ from \refeq{negfcurrent}. This is done by a midpoint integration scheme, using a regular integration grid of $N_E$ points in the range $\mu\pm 10k_BT$.

\section{Results}\label{results}

As an initial test of the Monte Carlo method we estimate the conductivity of nanowires. In this case of CdTe, with a square cross section of two by two unit cells. The conductivity is found by fitting against the conductances of wires, with lengths $10\times n$ unit cells, $n$ being in the range 1 to 10. In all calculations the temperature is set to 300 K, and the chemical potential is set to $\mu=4.5$ eV, which is about 1 eV above the conduction band edge.

The scattering models use B\"uttiker's approximation. Different values of $\gamma=\hbar/\tau$, namely 0.02, 0.04 and 0.06 eV were sampled. This corresponds to life times of $\tau=33$, 16 and 11 fs, respectively. The coupling to the leads is set to $\Gamma_l=0.2$ eV.

The scaling of the Monte Carlo method is compared to that of the direct and iterative schemes described in section \ref{direct_methods}. We required an accuracy of at least one percent for each method. In the Monte Carlo method this is assured by ending the sampling when a relative deviation of $10^{-2}$ is achieved. In the other methods, the accuracy is controlled by the resolution $N_E$ of the energy integration grid. We use a resolution of $N_E=13$ in all cases. This grid was chosen since it results in an error of one percent when compared to the converged $N_E\rightarrow\infty$ result in the $\gamma=0.06$ eV case.

\subsection{Transport results}

\begin{figure*}[tp]
    \centering
    \includegraphics[width=0.7\textwidth]{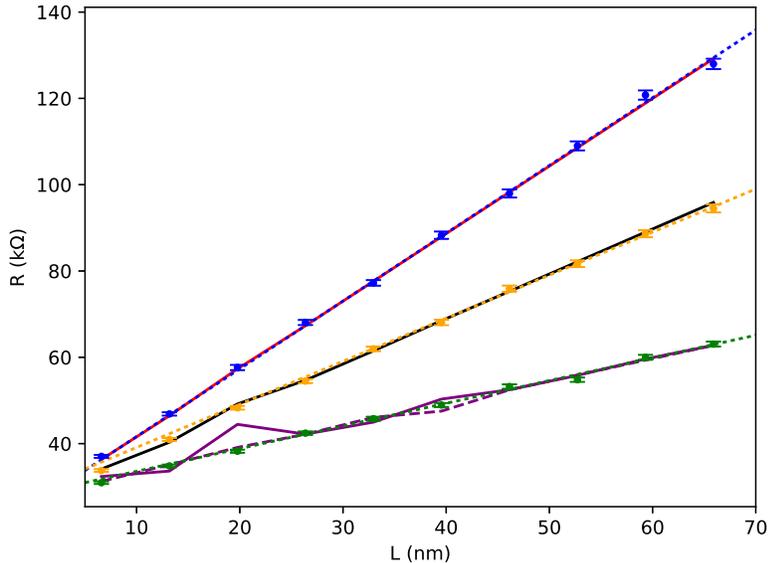}
\caption{Resistance $R=1/G$ of nanowires, as a function of wire length. Results of the iterative and direct methods are coinciding, and are shown as solid lines. Among these the results for $\gamma=0.06$, 0.04 and 0.02 eV are shown in red, black and purple, respectively. The Monte Carlo results are shown as error bars. Among these the results for $\gamma=0.06$, 0.04 and 0.02 eV are shown in blue, orange and green, respectively. We also show the iterative result for $\gamma=0.02$ eV with an extended integration grid of $N_E=26$. This is shown as a dashed purple line. The dotted lines represent linear functions fitted to the Monte Carlo results. These are shown in the same colors as the results to which they are fitted.}
\label{nanowire_conductance}
\end{figure*}

The wire resistance as a function of wire length is shown in figure \ref{nanowire_conductance}. The Monte Carlo calculations are consistent with the other results, except for $\gamma=0.02$ eV. We located this to be the result of $N_E=13$ being to coarse for this particular case. To show that integration is in fact the source of this discrepancy, we also include a calculation with $N_E=26$, where the results at $\gamma=0.02$ eV is indeed in better agreement.

The discrepancy between the $\gamma=0.02$ eV Monte Carlo and $N_E=13$ result is seen to be largest for short wire lengths. This indicates an increased sensitivity to the energy resolution due to edge effects penetrating further into the wires, which would be caused by the weaker scattering of $\gamma=0.02$ eV. For the longer wires we thus expect $N_E=13$ to be sufficient, and it is therefor used for the remaining analysis of the results.

As expected, the relation between resistance and wire length is close to linear. From 
figure \ref{nanowire_conductance} the resistivities of the wires can be estimated as the slope of linear fits. For the Monte Carlo method this yields 1570, 997.9 and 525.4 $\Omega$/nm for $\gamma=0.06$, 0.04 and 0.02 eV, respectively. The resistivities estimated from the direct/iterative solutions are correspondingly 1565, 1041 and 520.4 $\Omega$/nm.

\begin{figure*}[tp]
    \centering
    \begin{subfigure}[t]{0.49\textwidth}
        \centering
        \includegraphics[width=\textwidth]{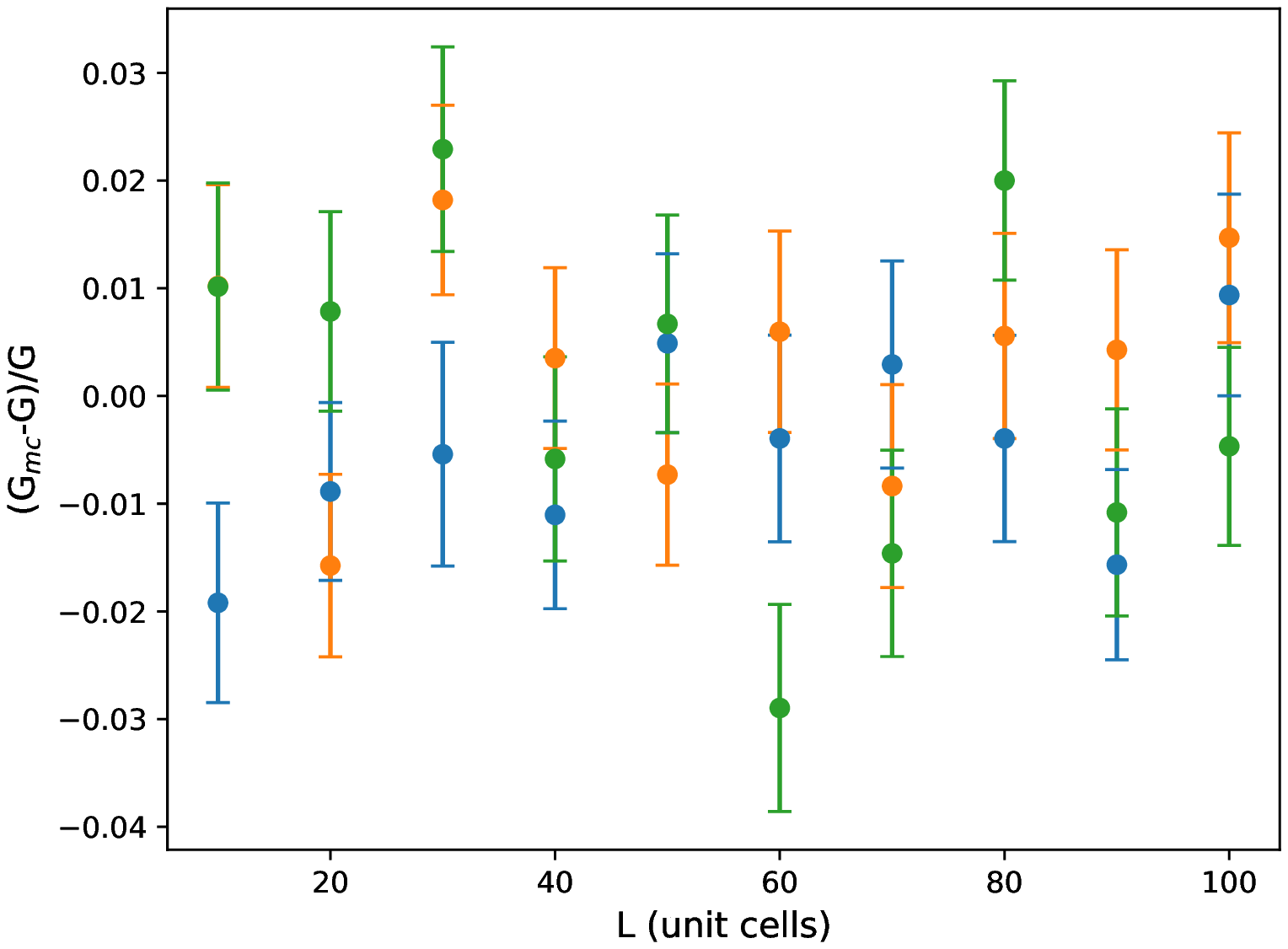}
        \caption{Relative error}
    \end{subfigure}
    \begin{subfigure}[t]{0.49\textwidth}
        \centering
        \includegraphics[width=\textwidth]{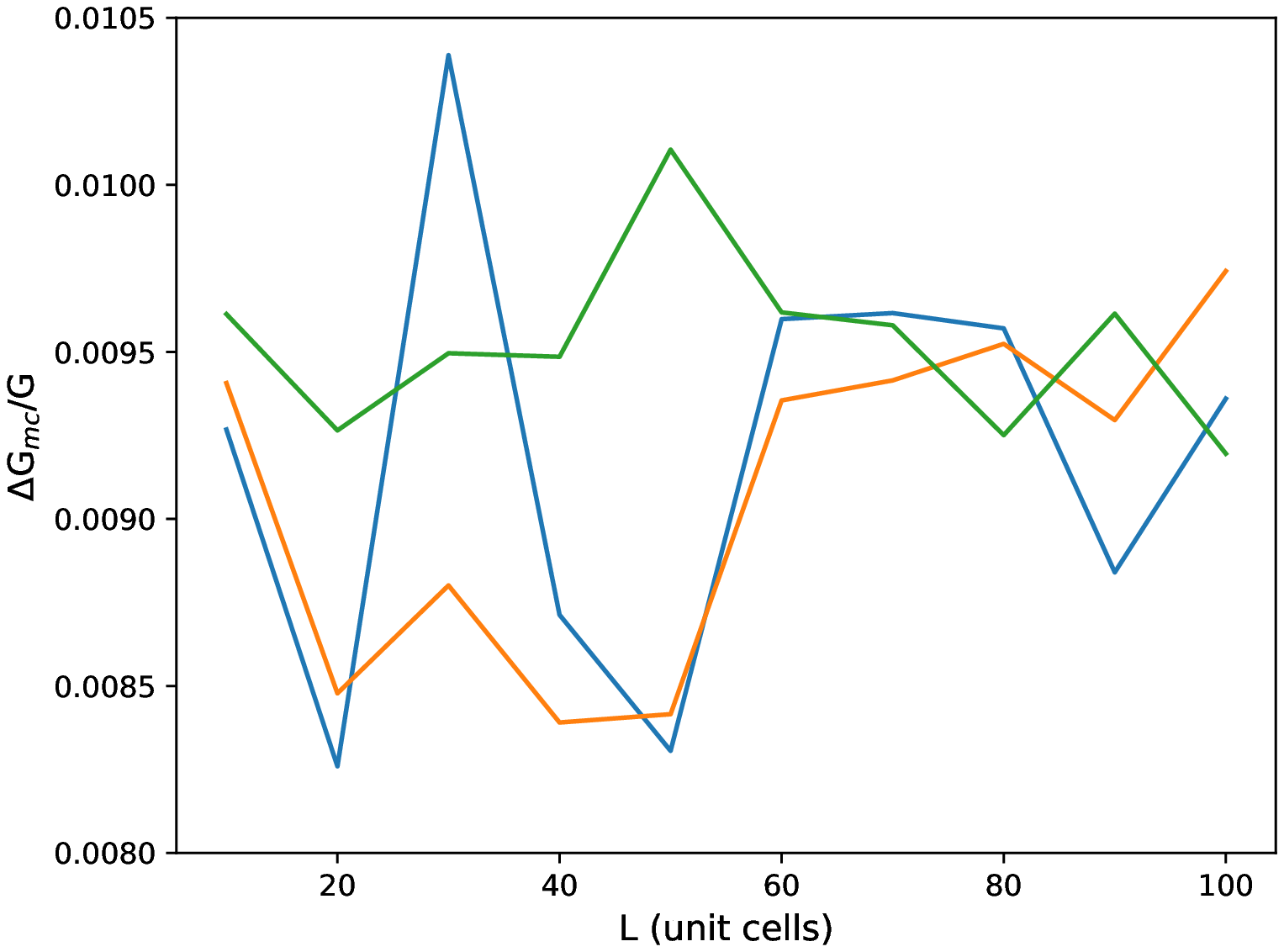}
        \caption{Relative deviation}
    \end{subfigure}
    \caption{Relative error and relative deviation as a function of wire length in unit cells. The relative error is calculated as $(G_{mc}-G)/G$, and the relative deviation as $\Delta G_{mc}/G$, where $\Delta G_{mc}$ is the standard deviation of the Monte Carlo result. The results are shown in blue, orange and green for $\gamma=0.06$, 0.04 and 0.02 eV, respectively.}
\label{nanowire_error}
\end{figure*}

The error bars in figure \ref{nanowire_conductance} are estimated as $\Delta R=\Delta(1/G)\approx\Delta G/G^2$. The relative deviation of $R$ is thus equal to that of $G$. The deviation is studied more closely in figure \ref{nanowire_error}, where we note that the relative deviation of the results are all approximately $0.01$ as was required. Next, we note that the relative error is always in the range $[-0.03,0.03]$. This means that the Monte Carlo results stay within three standard deviations from the direct/iterative solutions, as expected.

\subsection{Scaling}

\subsubsection{Direct and iterative method}

\begin{figure*}[tp]
    \centering
    \includegraphics[width=0.6\textwidth]{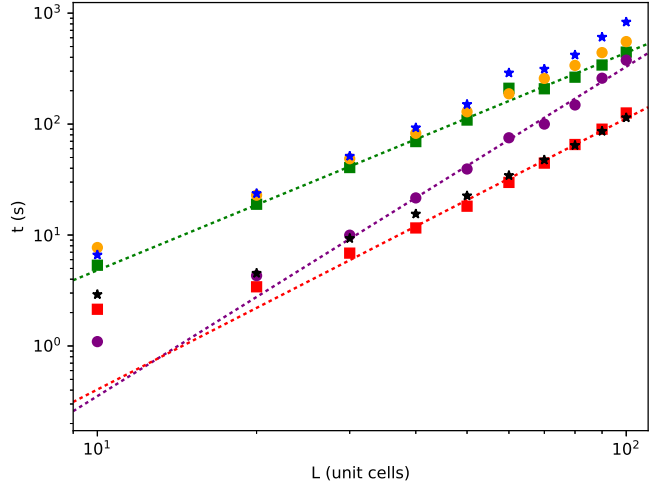}
    \caption{Computation time of the direct and iterative solution schemes. The blue stars and orange circles show the total computation time of the direct and iterative solutions, respectively. The green squares shows the time spent on executing the RGF algorithm, which coincides in the two methods. The purple circles and red squares show the time spent on solving \refeq{buteq}, using the the direct and the iterative method, respectively. These results are shown for $\gamma=0.06$ eV. Black stars show the computation time for the iterative solution using $\gamma=0.04$ eV. The dotted lines are fitting functions. These are shown in the same color as the results to which they are fitted.}
\label{nanowire_nonmcscaling}
\end{figure*}

In figure \ref{nanowire_nonmcscaling} we show the computation time of the direct and iterative methods described in section \ref{direct_methods}. The total computation time, the computation time devoted to the calculation of the Green's function $G^r$ using the RGF algorithm, and to the solution of \refeq{buteq} are shown. The total computation times of the direct and iterative methods are very similar. This is due to the fact that the computation time is dominated by the calculation of the Green's functions.

The computation time of the RGF algorithm, the direct solution of \refeq{buteq}, and the iterative solution of \refeq{buteq} are all fitted by power functions, which are also shown in figure \ref{nanowire_nonmcscaling}. The two first points are ignored in the fits, and the resulting fitting functions are respectively $t_{\mathrm{rgf}} = 0.05258\cdot L^{1.961}$, $t_{\mathrm{dir}} = 0.0003791\cdot L^{2.969}$ and $t_{\mathrm{itt}} = 0.001458\cdot L^{2.444}$.

From the fits we see that the Green's function calculation scales as $L^2$, while the direct solution of \refeq{buteq} scales as $L^3$, as expected. The iterative solution of \refeq{buteq} seems to scale close to $L^{\frac{5}{2}}$. The computation time of a single iteration should scale as $L^2$, so this result can either be interpreted as not reaching the asymptotic behavior, or that the number of iterations scales as $\sqrt{L}$. Adopting the latter interpretation, we conclude that for large $L$ the full direct solution method scales as $L^3$, while the full iterative method scales as $L^{\frac{5}{2}}$.

The operations required to calculate $G^r$, and for the direct solution of \refeq{buteq}, are independent of model parameters. Thus, their computation times will not be affected by the value of $\gamma$. In the case of the iterative method however, the value of $\gamma$ could in principle affect the required number of iterations. The computation times from the iterative method for $\gamma=0.04$ and $\gamma=0.06$ eV are also included in figure \ref{nanowire_nonmcscaling} to illustrate this. These are seen to coincide for large $L$, and we thus conclude that the scaling of the direct and iterative method is both unaffected by $\gamma$.

To summarize, we obtain the following fitting functions for the total computation time of the direct and iterative methods, respectively:
\begin{align}\label{nw_other_scaling}
t_{\mathrm{dir}} &= 0.04512\,\text{s}\cdot L^2+0.0003356\,\text{s}\cdot L^3,\,\,\text{and}\\\label{nw_itt_scaling}
t_{\mathrm{itt}} &= 0.04512\,\text{s}\cdot L^2+0.001162\,\text{s}\cdot L^{\frac{5}{2}}.
\end{align}
The coefficients of \refeq{nw_other_scaling} and \refeq{nw_itt_scaling}, which differ slightly from those of the fits in figure \ref{nanowire_nonmcscaling}, represent the fits obtained with fixed exponents.

\subsubsection{Monte Carlo method}\label{mcscalinganalysis}

\begin{figure*}[tp]
    \centering
    \begin{subfigure}[t]{0.49\textwidth}
        \centering
        \includegraphics[width=\textwidth]{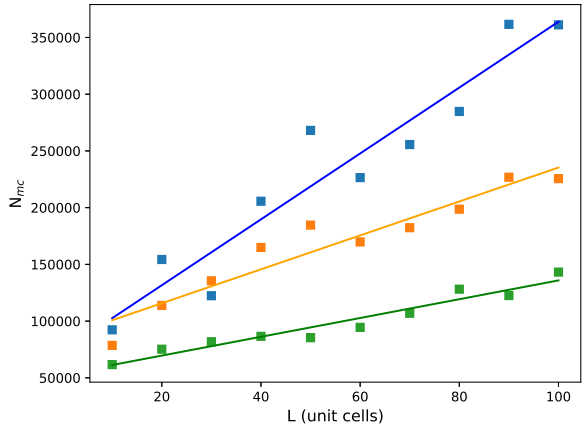}
        \caption{Number of samples}
	\label{nanowire_samples_fit}
    \end{subfigure}
    \begin{subfigure}[t]{0.49\textwidth}
        \centering
        \includegraphics[width=\textwidth]{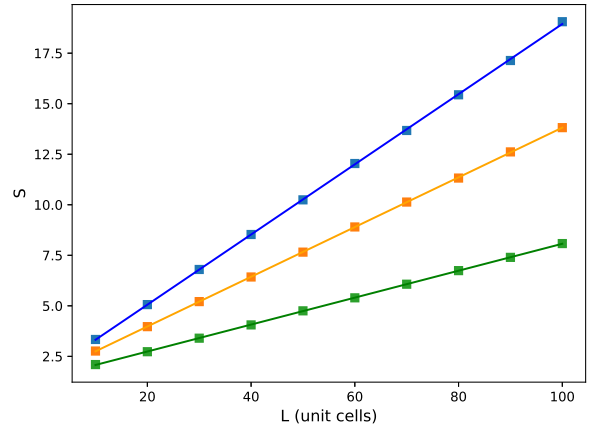}
        \caption{Scattering events per sample}
	\label{nanowire_steps_fit}
    \end{subfigure}
\caption{(a) Number of Monte Carlo samples, and (b) scattering events per sample. Results with $\gamma=0.06$, 0.04 and 0.02 eV are respectively shown as blue, orange and green squares. The results are fitted by linear functions. These are shown in the same colors as the results to which they are fitted.}
\label{nanowire_nfit}
\end{figure*}

Figure \ref{nanowire_nfit} shows the number of samples acquired during the Monte Carlo calculations. The figure also shows the average number of scattering events per sample. That is, the average number of executions of step 6' from section \ref{buttiker_section} before exiting at a lead. Linear fits to the results are also included. The fitting parameters are summarized in table \ref{nanowire_fitstable}.

\begin{table*}[tp]
\centering
\begin{tabular}{|l|l|l|l|l|l|l|}
\hline
$\gamma$ (eV) & $a_n$ & $b_n$ & $a_s$ & $b_s$ & $a_n/\gamma$ (eV$^{-1}$) & $a_s/\gamma$ (eV$^{-1}$) \\
\hline
0.06 & 2903. & 73542. & 0.1736 & 1.583 & 48383. & 2.893 \\
0.04 & 1495. & 85826. & 0.1230 & 1.518 & 37375. & 3.075 \\
0.02 & 828.2 & 53042. & 0.06663 & 1.407 & 41410. & 3.331 \\
\hline
\end{tabular}
\caption{Parameters of the linear fits shown in figure \ref{nanowire_nfit}. These model the number of samples $N_{mc}=a_nL+b_n$, and the average number of electron steps per sample $s=a_sL+b_s$.}
\label{nanowire_fitstable}
\end{table*}

We observe that both the number of samples, and the number of scattering events per sample, have a linear behavior. For the number of samples, this can be explained as follows: The number of samples required to obtain an accuracy of 0.01 will be $n=\Delta X^2/\langle X\rangle^2/0.01^2$, where $X$ is the estimator. $X$ is given by \refeq{elast_lin_xpq}, which with some approximation can be said to be binomially distributed, since most of the variation will originate from the delta function $\delta_{rq}$. In that case, $\langle X\rangle=p$ and $\Delta X^2=p(1-p)$, where the binomial probability $p$ is given by the transmission function $\mathcal{T}=(A+L/\lambda)^{-1}$. Putting this together we get $n=(A-1+L/\lambda)\cdot 10000$, which is a linear function in $L$.

The linear behavior of the number of steps per sample can also be understood. The average number of steps per sample can be expressed as
\begin{align}\label{avstepsint}
s=\int_0^L\mathrm{d}xp(x)s(x),
\end{align}
where $p(x)$ is the probability that the electron reaches a distance $x$ into the wire, and $s(x)$ is the average number of steps taken by an electron reaching this distance. The probability of reaching at least a distance $x$, is given by the transmission function $\mathcal{T}(x)$, which scales as $\lambda/x$. The probability $p(x)$ of an electron reaching a distance of exactly $x$ is then given by the derivative of $\mathcal{T}$, and is thus proportional to $\lambda/x^2$. Finally, the electron is undergoing a random diffusion type motion through the nanowire, which means the number of steps required to move a distance $x$ scales as $(x/\lambda)^2$. By inserting these estimates into \refeq{avstepsint} we get $s\sim L/\lambda$.

In fact, not only does the number of samples and the number of steps per sample show linear behavior, both are also proportional to $1/\lambda$. Since the mean free path $\lambda$ should be inversely proportional to the scattering rate $\gamma$, the slopes of the fitting functions in figure \ref{nanowire_nfit} should be proportional to $\gamma$. Table \ref{nanowire_fitstable} also shows the ratios $a/\gamma$, which is indeed seen to be roughly independent of $\gamma$. The average of all three ratios is respectively 42389 and 3.1 eV$^{-1}$ for the fits of figures \ref{nanowire_samples_fit} and \ref{nanowire_steps_fit}. The average constant term $b_n$ is found to be 70803. From this, we obtain
\begin{align}\label{nw_n_models}
N_{mc}&=42389\cdot \frac{\gamma L}{\text{eV}}+70803,\\\label{nw_s_model}
s&=3.1\cdot \frac{\gamma L}{\text{eV}},
\end{align}
which we assume is valid for all values of $\gamma$. We have here ignored the constant term $b_s$, since this is too small to be of significance at the length scales of these calculations.

\begin{figure*}[tp]
    \centering
    \includegraphics[width=0.6\textwidth]{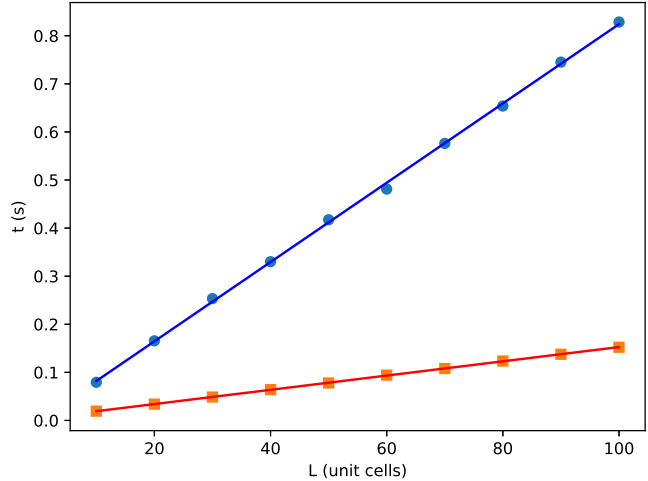}
    \caption{The time $t_d$ required for a single calculation of the diagonal of $G^r$ is shown as blue circles, while the time $t_s$ required for the execution of a single electron step is shown as orange squares. Only the calculation with $\gamma=0.06$ eV is included. The other values of $\gamma$ give nearly identical results and is not shown. The solid lines represent linear fits. These are in the same color as the results to which they are fitted.}
\label{nanowire_rgftime}
\end{figure*}

The computation time of the Monte Carlo method depend on $N_{mc}$ and $s$, but also on the average time $t_s$ spent per scattering event, and on the average time $t_d$ required to calculate the diagonal of $G^r$, performed once per Monte Carlo sample. Estimates of these quantities obtained from our calculations are shown in figure \ref{nanowire_rgftime}. The figure also contains linear fits. These are $t_d=0.00824\,\text{s}\cdot L -0.000445$ s, and $t_s = 0.00148\,\text{s}\cdot L + 0.00438$ s. The fits closely resemble the calculations, showing that these computation times have close to perfect linear scaling. This is reasonable since the most time consuming operation per scattering event is the calculation of a single column of $G^r$. The calculation of single columns as well as the diagonal of $G^r$ are linearly scaling operations in the RGF algorithm.

We adopt the following models for the computation times $t_d$ and $t_s$:
\begin{align}\label{nw_time_models}
t_d&=0.00824\,\text{s}\cdot L,\\\label{nw_time_models_s}
t_s&=0.00148\,\text{s}\cdot L,
\end{align}
where the constant terms have been ignored due to their small size. The computation time required for a single Monte Carlo sample will be $t_d+st_s$. Thus, combining the expressions \refeq{nw_n_models} to \refeq{nw_time_models_s}, we obtain the model
\begin{align}\label{nw_mc_scaling}
t&=\left(t_d+s\,t_s\right)N_{mc}=\left(0.00824\,\text{s}\cdot L+3.1\cdot \frac{\gamma L}{\text{eV}}\cdot 0.00148\,\text{s}\cdot L\right)\cdot\left(42389\cdot \frac{\gamma L}{\text{eV}}+70803\right)
\end{align}
for the total computation time.

\subsubsection{Comparison between the methods}

\begin{figure*}[tp]
    \centering
    \includegraphics[width=0.7\textwidth]{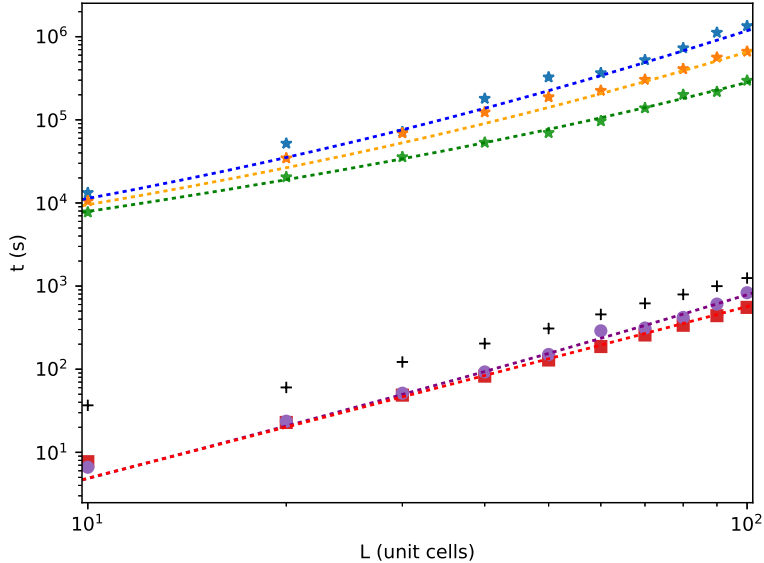}
    \caption{Total computation time. The purple circles and red squares represent the direct solution and the iterative method, respectively. These are compared to their scaling models given by \refeq{nw_other_scaling} and \refeq{nw_itt_scaling}, which are shown as dotted lines in the same colors. The black crosses show the iterative results where an energy integration grid of $N_E=26$ rather than $N_E=13$ points is used. The blue, orange and green stars represent Monte Carlo calculations with $\gamma=0.06$, 0.04 and 0.02 eV, respectively. These are compared to the scaling model given by \refeq{nw_mc_scaling}, which are shown as dotted lines in the same colors as the corresponding results.}
\label{nanowire_alltimes}
\end{figure*}

The total computation times of all methods are shown in figure \ref{nanowire_alltimes}, where they are compared to their respective fitted models. These models are given by \refeq{nw_mc_scaling} in the case of the Monte Carlo calculations, and respectively by \refeq{nw_other_scaling} and \refeq{nw_itt_scaling} in the case of the direct and iterative approaches. The computation time of the Monte Carlo method is roughly three orders of magnitude larger than that of the standard approaches. However, since the iterative method scales as $L^{\frac{5}{2}}$, direct solution as $L^3$, and the Monte Carlo method as $L(L/\lambda)^2$, the Monte Carlo method will in fact be faster if the mean free path $\lambda$ is sufficiently large.

\begin{figure*}[tp]
    \centering
    \includegraphics[width=0.7\textwidth]{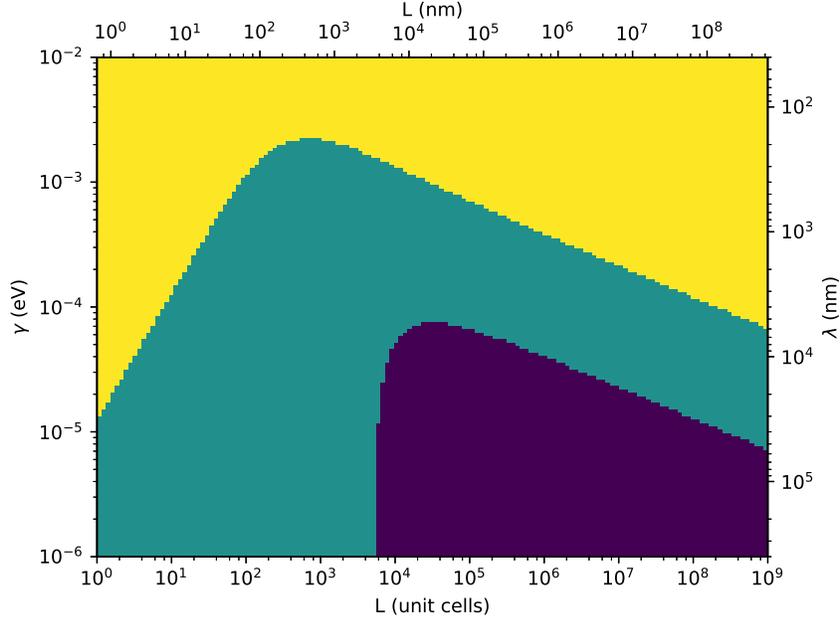}
    \caption{Comparison of the predicted computation times of the Monte Carlo and iterative methods. The dark purple region shows where the Monte Carlo methods is faster than the iterative method. The iterative method is faster than our prototype implementation of the Monte Carlo method in both the yellow and the green region, but in the green region the Monte Carlo method is the fastest when combined with the interpolating technique discussed in section \ref{interpolation_discuss}.}
\label{nanowire_scaling}
\end{figure*}

Figure \ref{nanowire_scaling} shows the parameter region where the Monte Carlo method is faster than the iterative method, as predicted by the models \refeq{nw_mc_scaling} and \refeq{nw_itt_scaling}. The figure shows that $\gamma$ must be smaller than 70 $\mu$eV in order for the Monte Carlo method to be faster for any value of $L$. Figure \ref{nanowire_scaling} also makes reference to a third method, which will be discussed in section \ref{interpolation_discuss}.

By the arguments above, the relevant parameter determining the required computation time of the Monte Carlo method is actually the mean free path $\lambda$, rather than the scattering strength $\gamma$. We can obtain some very rough estimates of the mean free paths in our calculations by utilizing the fact that edge effects will penetrate roughly a few mean free paths into the wires. Observing figure \ref{nanowire_conductance}, we see that in the $\gamma=0.04$ eV case, edge effects are causing a deviance from the linear behavior for wire lengths up to $20$ nm, while the same edge effects cause deviations at wire lengths of 40 nm in the $\gamma=0.02$ eV case.

Assuming that edge effects are visible for wire lengths shorter than $\sim 2\lambda$, we estimate the mean free paths in those two cases to be 10 and 20 nm. Since $\lambda$ should be inversely proportional to the scattering rates, we obtain the model $\lambda=0.4\,\text{eVnm}/\gamma$. This would mean the charge carriers in our example nanowire have an approximate average group velocity of 4 eV$\text{\AA}$, which is a reasonable number. Adopting this model, the condition $\gamma<70\,\mu$eV can be translated to the requirement that $\lambda$ must be larger than approximately $6\,\mu$m. This is unlikely to be the case in most practical calculations.

\section{Discussion}\label{discussion}

\subsection{Performance}\label{performance}

Since the Monte Carlo method requires large mean free paths $>$ 6 $\mu$m in order to be faster than the iterative method, our current implementation is not useful for the chosen example. This situation is unlikely to change for other finite size geometries. Thus, in order for the method to be useful, we are in need of an improved implementation. There are multiple ways in which both the method itself, and its implementation can be improved, performance wise. Some of these will be discussed in the following. Their implementation and verification will be postponed to future studies.

\subsubsection{Compiled solvers}

Our prototype tests make use of a straightforward Python implementation of the Monte Carlo solver. It is well known that there is some overhead, possibly orders of magnitude due to the non-compiled nature of Python. It is important to emphasize that also the other methods in our tests were implemented in Python. However, most of these made use of precompiled solvers, such as $gmres$ in the case of the iterative method. Thus, a direct comparison between the methods is not entirely fair, and we can expect improved performance by writing the Monte Carlo solver in a compiled language.

However, a significant improvement is not guarantied to follow from such conversion alone. While the RGF algorithm was also implemented in Python, it utilizes a large number of matrix inversions and products, functionality which is already precompiled through NumPy. Any expected improvements by switching the entire method to a compiled implementation will be limited by the ratio in cost between these matrix operations and the Python overhead. This will largely depend on the size of the matrices, which in the case tested here was only 32 by 32.

\subsubsection{Systems with k-dependence}

The relative difference in performance between the Monte Carlo method and the other approaches can be expected to improve in systems with a k-dependence. If the nanowires of section \ref{results} were replaced with thin-films of varying thickness, the Hamiltonian and other operators would then have a two dimensional k-dependence. If a $10\times 10$ k-point grid was used for the Brillouin zone, then the number of unknowns in \refeq{buteq} would increase by a factor of 100. Accordingly, the direct and iterative solution schemes would respectively be slower by six and four orders of magnitude. On the other hand, the change in computation time of the Monte Carlo method would be determined by the change in the variance of the estimator. Although the variance can be expected to increase, a change by four orders of magnitude is unlikely.

\subsubsection{Parallelization}

One large benefit to Monte Carlo calculations, is that they are very easy to parallelize, and that they scale near perfectly with processor distribution. Thus, although what we refer to as the total computation time would not be reduced, the real time requirement of the calculations could be significantly improved by utilizing high throughput parallelization. In fact, we have already made use of parallelization in our test calculations, which were executed in parallel using 64 CPUs. The time shown in figure \ref{nanowire_alltimes} is the total computation time, and the real execution times of the Monte Carlo calculations are thus in reality lower by a factor of 64.

While the iterative and direct methods can also be parallelized, this is less efficient when the number of processors exceeds the number of integration points. The efficiency of a parallelized Monte Carlo calculations could possibly be increased even further by making use of Graphics Processing Units (GPUs). The use of GPUs requires that the memory requirements of each process are very small. Such small memory requirement is in fact one of the benefits of the Monte Carlo method.

\subsubsection{Wave function based solver}

Switching to a wave function based solver is another possibility. That is, instead of calculating $G^r$ explicitly, the wave functions $|\psi_n\rangle=G^r|i_n\rangle$ is calculated by solving \refeq{modified_schrodinger}. In the calculations of section \ref{results} a wave function based solver would not cause a very significant improvement. This is due to the utilization of the RGF algorithm, which scales linearly with system size along one spatial dimension. However, for the other spatial dimensions the RGF algorithm has cubic scaling. It is thus best suited for long structures, such as thin nanowires.

In more bulk like systems, wave function based methods utilizing efficient sparse linear solvers, such as those employed by the ballistic transport software Kwant\cite{kwant} are most likely better suited. Alternatively, iterative sparse solvers like the $gmres$ method would be less memory demanding, and thus better suited for parallelization.

\subsubsection{Interpolation}\label{interpolation_discuss}

The most time consuming part of the Monte Carlo method of section \ref{buttiker_section}, is the calculation of the wave functions $|\psi_n\rangle$. These are primarily needed in the calculation of the probabilities $p_i$ of \refeq{draw_i_simult}. Thus, we could speed up the calculations by interpolating these probabilities. Specifically, we tested this approach by checking whether two probabilities $p_i$ calculated at nearby energies differ by less than five percent. If this is the case, then all values of $p_i$ at energies between these two points are found by linear interpolation. This approach resulted in an order of magnitude reduction in computational time without any visible loss in accuracy of the conductance estimates.

In figure \ref{nanowire_scaling} we also show the parameter range in which this interpolating Monte Carlo method is faster than the iterative method. The scaling estimate of the interpolating method was obtained using an analysis similar to that of section \ref{mcscalinganalysis}. Figure \ref{nanowire_scaling} shows that the minimal mean free path requirement is reduced to about 200 nm when the interpolating method is used. This is a significant improvement, but still not quite in the range of practical applicability. However, because the Monte Carlo method consists only of very simple operations when the wave function calculations are eliminated, we believe that the computational time of the interpolating method is mainly determined by the Python overhead. We expect further reduction in efficiency when traversing this solution to a compiled language.

\subsubsection{Improved estimator}\label{estimator_discuss}

Finally, the efficiency of the Monte Carlo method could also be improved by reducing the variance of the estimators. One way of approaching this is the use of weighted Monte Carlo methods, like those applied in Boltzmann calculations\cite{jacoboni}. This would involve modifying the probability distributions of section \ref{intuitive_integral} to new probabilities $p'$, and then weighting the estimator with the probability ratios $p/p'$. If this is done correctly, an estimator $X'$ will have smaller variance. Less samples will thus be required to achieve a particular accuracy.

There is also another way in which the variance can be reduced for bulk quantities like the electrical conductivity. From \refeq{nw_n_models} we see that the number of samples $N_{mc}$ is increased significantly by the presence of a large constant term. This term stems from the constant term $A$ in $1/\mathcal{T}=A+L/\lambda$, which reduces the binomial probability and increases the variance. The term $A$ is related to the probability of transmission between the system and the leads. Thus, we could reduce the variance of the estimator by eliminating the leads from the expression. Intuitively, it should be possible to estimate conductivity from the distribution of electron step lengths $|x_{n+1}-x_n|$, rather than from expressions like $X_{pq}$, which involve the leads $p$ and $q$. Here $x_n=\langle\psi_n|x|\psi_n\rangle$.

\subsection{Generalizations}

In addition to the suggested strategies to improve the computational efficiency of the calculations, it is also possible to incorporate additional physics in the Monte Carlo model.

\subsubsection{Generalizations of the scattering model}

The test calculations of this work utilize two approximations to the scattering model: 1) B\"uttiker's approximation, and 2) elastic scattering. In calculations intended to model material systems for technology applications, these approximations are likely to be too limiting. In particular, self energies derived from various perturbative schemes, e.g. modeling electron-phonon interaction and disorder scattering will rarely conform to B\"uttiker's approximation.

If such models are employed, it is necessary to use the more general method of section \ref{intuitive_integral}, rather than the simplified algorithm of section \ref{buttiker_section}. This requires diagonalization of the operators $\Lambda^{(n+1)}$, which increases the memory requirements and the computation time of the method. However, since scattering is usually a local phenomenon, $\Lambda^{(n+1)}$ can most likely be approximated as a sparse matrix, possibly ordered, so that the increased computational requirements are manageable.

In addition, both electron-electron and electron-phonon interactions give rise to inelastic scattering. While acoustic phonon scattering can often be approximated as elastic, this is not the case with scattering of optical phonons. Also, in applications where energy relaxation, for instance thermalization of states is important, even acoustic scattering may need to be modeled as inelastic. In high field applications this is merely a question of finding a way to draw energies from the probability distributions of \refeq{mc_directe_pofe} or \refeq{draw_E_last}, followed by application of the algorithms in sections \ref{intuitive_integral} or \ref{buttiker_section}.

However, when one is interested in linear transport coefficients such as $G_{pq}$, this approach is not well suited. At low fields the currents $I_q$ will be very small, so that the signal to noise ratio of the estimators $X_q$ will also be low. Thus, in the linear limit we believe a better approach is to find an expression which generalizes \refeq{elast_lin_xpq} to inelastic scattering. This generalized expression is likely to be considerably more involved.

\subsubsection{Generalizations of the Hamiltonian}

The Hamiltonian employed in our calculations could also be generalized in various ways. Within a tight binding framework, one could improve the model either by including more orbitals per atom, or by including interactions of longer range than the nearest neighbor. In addition, one could also include effects such as a Poisson potential or strain dependent model parameters, and interface the Hamiltonian directly to some ab-initio software, employing for instance Density Functional Theory (DFT) or post DFT methods.

\subsubsection{Nonlinearities}\label{discuss_hf}

The Monte Carlo methods described in sections \ref{general_integral}, \ref{intuitive_integral} and \ref{buttiker_section} are all based on the assumption that the function $\Xi(E,E',G)$ defined in \refeq{xiint_defined} is linear. There are two ways in which this linearity assumption could fail: 1) The scattering function $\Lambda(E,E',G)$ could itself be nonlinear, or 2) the Green's functions $G^r$ and $G^a$ could be functions of $G^<$.

As discussed in appendix \ref{apdix_lin}, $\Lambda$ will typically be nonlinear if one makes use of perturbation theory to orders higher than one, or if one includes electron-electron scattering to any order. Also, as discussed in section \ref{lin_limit_section}, $G^r$ and $G^a$ can be functions of $G^<$ either through the scattering mechanism, or for instance a Poisson potential. Thus, the function $\Xi(E,E',G)$ will in most realistic cases be nonlinear in $G$.

We will handle the nonlinearity by introducing an object $x=F(G^<)$, and write the function $\Xi$ as $\Xi(E,E',G, x)$, where $\Xi$ is now linear in the third argument. Note that this is always possible in principle, since we could choose $x=G^<$. The reason for using the more general expression is that it will sometimes be more efficient to choose $x$ differently. For instance, if the nonlinearity comes from a Poisson potential, one could choose $x=\rho$, where $\rho$ is the charge distribution.

Given any fixed value of $x$, $\Xi$ is linear, and we can solve \refeq{g_less_exp} using the Monte Carlo method. The solution is a lesser Green's function $\tilde{G}^<(x)$ which depends on the value used for $x$. $\tilde{G}^<(x)$ will only equal the true solution $G^<$ if $x$ is set to its true value $x^\star$. That is, $G^<=\tilde{G}^<(x^\star)$. Since $x^\star=F(G^<)$, we can find $x^\star$ by solving the equation $x=F(\tilde{G}^<(x))$. In particular, if $F(\tilde{G}^<(x))$ can be estimated in the Monte Carlo method using some estimator $X$, then $x^\star$ is the solution of $\langle X|x=x^\star\rangle=x^\star$. Here the notation $X|x=x'$ means that the estimator is calculated in a simulation where $x=x'$. 

One way of solving this equation is to construct a sequence of random estimates $X_n=X|x=x_n$, where $x_n$ is always the average of all previous estimates. That is, 
\begin{align}
x_n=\frac{1}{n}\sum_{k=0}^{n-1}X_k
\end{align}
for $k>0$, while $x_0$ equals some initial guess. Intuitively, $x_n$ should approach the correct value of $x$ as $n\rightarrow\infty$. Such an approach is sometimes employed when Monte Carlo simulations are used to solve the Boltzmann equation in the presence of a Poisson potential. It is referred to as the incident flux approach\cite{lundstrom}.

\subsubsection{Generalization to phonons}

In principle, the Monte Carlo method described here could also be applied to phonon transport. Since equations similar to \refeq{eq_for_gr} and \refeq{eq_for_gl} can be found also in phonon NEGF formalism, most of the expressions in section \ref{theory} should carry directly over to the phonon method. An important exception however, is assumption (iv), which is a consequence of charge conservation. Unless a similar conservation condition can be shown to hold also for phonons, Monte Carlo estimates must be based on \refeq{X_def_intmc} rather than \refeq{X_def_intmc_cc}.

\section{Conclusions}\label{conclusion}

We have shown that a Monte Carlo scheme similar to that used for the solution of the Boltzmann equation can also be used to solve the NEGF equations, as long as certain assumptions about the scattering self energies are satisfied. While the most general Monte Carlo method, described in section \ref{general_integral}, requires only assumptions (i) and (ii) to be satisfied, the more specialized method described in section \ref{intuitive_integral} requires also assumption (iii). In addition, the simplified estimator of \refeq{X_def_intmc_cc} is based upon assumption (iv).

The described Monte Carlo scheme solves \refeq{eq_for_gl}, the NEGF equation for $G^<$. In the case of elastic scattering in the linear limit, we have shown that we can use the equilibrium value of $G^r$, so that \refeq{eq_for_gr} need not be solved self consistently with \refeq{eq_for_gl}. Possibly, some generalization of this can be found also for inelastic scattering. In nonlinear transport problems, we suggest combining the Monte Carlo method with an iterative scheme, similar to what is done when Monte Carlo simulations are used to solve the Boltzmann equation in the presence of a Poisson potential.

We tested the method with a scattering model conforming to B\"uttiker's approximation. The tests show that at the level of accuracy required, the Monte Carlo method is faster than the alternative approaches in a specific regime. However, this regime involves mean free paths larger than what can be found in most applications. Our prototype implementation was performed to demonstrate the concept and present introductory performance numbers to guide future studies. As we have discussed, there are several ways in which the current implementation can be be improved to reduce the computation time. More work is thus needed to demonstrate how successful such improvements will be, and whether the discussed physical generalizations are applicable in practice.

\appendix

\section{Justification for the scattering assumptions}\label{apdix_assump}

In this appendix, we try to provide some justification for the assumptions (i) - (v) introduced in the main text. A formal proof that these assumptions are always valid, is not the scope of this appendix, and this may not even be the case. Instead, we propose heuristic arguments, in an attempt to justify why these assumptions are reasonable, or at least why they are reasonable as approximations.

\subsection{Assumption (i), Linearity}\label{apdix_lin}

In the case of electron-phonon scattering in the self consistent Born approximation, the self energy $\Sigma^<_s$ can be expressed as\cite{datta, jacoboni, lake1997single}
\begin{align}\label{electron_phonon}
\Sigma^<_{sij}(E)=\sum_{kl}\int\mathrm{dE'}\,D^{ij}_{kl}(E')G^<_{kl}(E-E').
\end{align}
This expression can be written on the form of \refeq{lesser_scatt_express}, with 
\begin{align}\label{electron_phonon_lambda}
\Lambda(E,E',G)_{ij}=\sum_{kl}D^{ij}_{kl}(E-E')G_{kl},
\end{align}
which is seen to be linear in $G$ as required. Similar expressions apply for scattering on impurities and other forms of disorder, and indeed for self consistent Born approximations of scattering on any other field than the electron field itself.

In the case of electron-electron scattering, the lowest order perturbation would instead take the form
\begin{align}\label{electron-electron}
\Sigma^<_{sij}(E)=\sum_{klmn}\int\mathrm{dE'}\,F_{ij}^{klmn}(E')G^<_{kl}(E-E')G^<_{mn}(E'),
\end{align}
and $\Lambda$ would thus be second order in $G$. Clearly, other scattering mechanisms will also result in nonlinear $\Lambda$ if higher order expansions than \refeq{electron_phonon} are used. However, 
in principle we can rewrite $\Lambda$ as $\Lambda(E,E',G,G^<)$, where $\Lambda$ is still linear in the third argument. For instance, \refeq{electron-electron} could be written on the form of \refeq{lesser_scatt_express} with 
\begin{align}
\Lambda(E,E',G, G^<)_{ij}=\sum_{klmn}F_{ij}^{klmn}(E')G^<_{kl}(E-E')G_{mn},
\end{align}
which is now linear in $G$. However, this is only useful if the fourth argument of $\Lambda$ can be approximated in some other way. One way of doing this is to expand it in terms of the unperturbed Green's function $g^<$. Alternatively, one could try the approach suggested in section \ref{discuss_hf}.

\subsection{Assumption (ii), Hermiticity}\label{apdix_herm}

Both the Green's function $G^<(E)$ and the self energy $\Sigma^<(E)$ are anti-Hermitian operators\cite{datta}. Since the lead contributions to \refeq{sigmas_comp_l} are Hermitian, so must $\Sigma^<_s$ be. But this just means that the integral in \refeq{lesser_scatt_express} is anti-Hermitian, and does not necessarily mean that $\Lambda(E,E',G)$ is Hermitian for arbitrary $E'$ and $G$. However, the only physical content of the function $\Lambda$ is in fact the self energy resulting from \refeq{lesser_scatt_express}. Thus, we are free to arbitrarily modify this function, as long as $\Sigma^<_s$ is preserved. So if $\Lambda$ does not conform to assumption (ii), we can simply define a new function $\Lambda'$ as 
\begin{align}
\Lambda'(E,E',G)=\frac{\Lambda(E,E',G)+\Lambda(E,E',G^\dag)^\dag}{2}.
\end{align}
We then have
\begin{align}
\int\mathrm{d}E'\Lambda'(E,E',G^<(E'))&=\frac{1}{2}\int\mathrm{d}E'\Lambda(E,E',G^<(E'))+\frac{1}{2}\int\mathrm{d}E'\Lambda(E,E',-G^<(E'))^\dag\\\nonumber
&=\frac{\Sigma^<_s(E)-\Sigma^<_s(E)^\dag}{2}=\Sigma^<_s(E),
\end{align}
as required. In addition, it is easily seen that if $G$ is Hermitian, then so is $\Lambda'(E,E',G)$, and that if $\Lambda$ is linear in $G$, then so is $\Lambda'$.

\subsection{Assumption (iii), Positivity}\label{apdix_pos}

We will first show that assumption (iii) holds for disorder scattering in the self consistent Born approximation. The self energy is then given by \refeq{electron_phonon}, where $D^{ij}_{kl}$ is given by
\begin{align}\label{d_of_dissorder}
D^{ij}_{kl}(E)=\left\langle V_{ik} V_{lj}\right\rangle\delta(E),
\end{align}
where $V$ is the disorder potential, and the brackets denote an ensemble average. Inserting this in \refeq{electron_phonon_lambda}, we find
\begin{align}
\langle\psi|\Lambda(E,E',G)|\psi\rangle=\sum_{klij}\psi^\star_i\psi_j\left\langle V_{ik} V_{lj}\right\rangle G_{kl}\delta(E-E')=\left\langle \langle\psi|VGV|\psi\rangle \right\rangle \delta(E-E').
\end{align}
Since $V$ is Hermitian, $\langle\psi|VGV|\psi\rangle$ is positive if $G$ is positively definite. Thus, so is $\langle\psi|\Lambda(E,E',G)|\psi\rangle$, and accordingly assumption (iii) is satisfied.

Next, will show that assumption (iii) holds for electron-phonon scattering in the self consistent Born approximation. The self energy is still given by \refeq{electron_phonon}, but the function $D^{ij}_{kl}(E)$ is now
\begin{align}\label{d_of_electron_phonon}
D^{ij}_{kl}(E)=\frac{1}{\hbar}\sum_{rs}U^r_{ik} U^s_{lj}D_{rs}(E),
\end{align}
where the potentials $U^r_{ij}$ describe perturbations to the electronic Hamiltonian introduced by lattice perturbations $x_r$, and
\begin{align}\label{drse_defined}
D_{rs}(E)=\int\mathrm{d}t\,\left\langle x_r(t) x_s(t')\right\rangle e^{iE(t-t')/\hbar},
\end{align}
where the brackets now denotes an average over the lattice state. \refeq{drse_defined} makes use of the assumption of a stationary process, so that $\left\langle x_r(t) x_s(t')\right\rangle$ is a function of $t-t'$ alone.

The first step is to show that the matrix $D(E)$ with elements given by \refeq{drse_defined} is positively definite for all $E$. Consider a set of functions $f_r(t)$. We have
\begin{align}\label{aa3_geq_integral}
\sum_{rs}\int\mathrm{d}t\int\mathrm{d}t'\,\left\langle x_r(t) x_s(t')\right\rangle f_r(t)f_s^\star(t')=\left\langle\left|\sum_{r}\int\mathrm{d}t\, x_r(t) f_r(t)\right|^2\right\rangle\geq 0.
\end{align}
We can also express the integral in the transformed variables $T=(t+t')/2$ and $\Delta=t'-t$. Making use of the fact that $\left\langle x_r(t) x_s(t')\right\rangle$ is a function of $\Delta$ alone, we get
\begin{align}\label{delta_transform}
&\sum_{rs}\int\mathrm{d}t\int\mathrm{d}t'\left\langle x_r(t) x_s(t')\right\rangle f_r(t)f^\star_s(t')\\\nonumber
=&\sum_{rs}\int\mathrm{d}\Delta\left\langle x_r(t) x_s(t')\right\rangle \int\mathrm{d}Tf_r\left(T-\frac{\Delta}{2}\right)f_s^\star\left(T+\frac{\Delta}{2}\right)
\end{align}
Now introducing the Fourier transforms $\phi_r(\omega)$ of the functions $f_r$, we can write
\begin{align}
&\int\mathrm{d}Tf_r\left(T-\frac{\Delta}{2}\right)f_s^\star\left(T+\frac{\Delta}{2}\right)=\frac{1}{(2\pi)^2}\int\mathrm{d}T\int\mathrm{d}\omega\int\mathrm{d}\omega'\phi_r(\omega)\phi_s^\star(\omega')e^{-i\omega(T-\Delta/2)+i\omega'(T+\Delta/2)}\\\nonumber
&=\frac{1}{2\pi}\int\mathrm{d}\omega\int\mathrm{d}\omega'\phi_r(\omega)\phi_s^\star(\omega')e^{i\omega\Delta/2+i\omega'\Delta/2}\delta(\omega-\omega')
=\frac{1}{2\pi}\int\mathrm{d}\omega\,\phi_r(\omega)\phi_s^\star(\omega)e^{i\omega\Delta}.
\end{align}
Inserting this in \refeq{delta_transform}, we get
\begin{align}\label{aa3_d_integral}
\sum_{rs}\int\mathrm{d}t\int\mathrm{d}t'\left\langle x_r(t) x_s^\star(t')\right\rangle f_r(t)f_s(t')=&\frac{1}{2\pi}\sum_{rs}\int\mathrm{d}\Delta\int\mathrm{d}\omega\,\left\langle x_r(t) x_s(t')\right\rangle \phi_r(\omega)\phi_s^\star(\omega)e^{i\omega\Delta}\\\nonumber
=&\frac{1}{2\pi}\sum_{rs}\int\mathrm{d}\omega\phi_r(\omega)\phi_s^\star(\omega)D_{rs}(\omega/\hbar).
\end{align}
Let us now specify $f_r(t)=v_rf(t)$, so that $\phi_r(\omega)=v_r\phi(\omega)$ for some vector $\boldsymbol{v}=[v_r]$. Comparing \refeq{aa3_d_integral} to \refeq{aa3_geq_integral}, we see that
\begin{align}\label{aa3_lasd_d_stuff}
\sum_{rs}\int\mathrm{d}\omega\,|\phi(\omega)|^2v_rv_s^\star D_{rs}(\omega/\hbar)=\int\mathrm{d}\omega\,|\phi(\omega)|^2\boldsymbol{v}^\dag D(\omega/\hbar)\boldsymbol{v}\geq 0.
\end{align}
Since this must hold for arbitrary functions $f$, and arbitrary vectors $\boldsymbol{v}$, the matrix $D(E)$ must be positively definite for all $E$.

We can reexpress \refeq{d_of_electron_phonon} in terms of any basis of phonon modes $\tilde{x}_\rho$. Thus, let us use a basis which diagonalizes $D(E)$. \refeq{d_of_electron_phonon} then becomes
\begin{align}\label{phonon_d_diagonal}
D^{ij}_{kl}(E)=\frac{1}{\hbar}\sum_{\rho}\tilde{U}^\rho_{ik} \tilde{U}^\rho_{lj}\lambda_\rho(E),
\end{align}
where the eigenvalues $\lambda_\rho(E)$ of $D(E)$ are all positive. Inserting \refeq{phonon_d_diagonal} in \refeq{electron_phonon_lambda}, we find
\begin{align}
\langle\psi|\Lambda(E,E',G)|\psi\rangle=\frac{1}{\hbar}\sum_{klij\rho}\psi^\star_i\psi_j\tilde{U}^\rho_{ik} \tilde{U}^\rho_{lj}\lambda_\rho(E-E') G_{kl}=\frac{1}{\hbar}\sum_\rho \langle\psi|\tilde{U}^\rho G\tilde{U}^\rho|\psi\rangle \lambda_\rho(E-E')
\end{align}
Since the potentials $\tilde{U}^\rho$ are again Hermitian, $\langle\psi|\tilde{U}^\rho G\tilde{U}^\rho|\psi\rangle$ is positive if $G$ is positively definite. Thus, so is $\langle\psi|\Lambda(E,E',G)|\psi\rangle$, and accordingly assumption (iii) is satisfied also for phonon scattering.

This means that under the assumption of a stationary process, assumption (iii) is at least satisfied for two very common scattering models, namely self consistent Born approximations of disorder and phonon scattering. Generalizing to higher orders is more difficult, but we can make a heuristic argument. By an argument very similar to that of \refeq{aa3_geq_integral} to \refeq{aa3_lasd_d_stuff}, one can also show that $-iG^<(E)$ is positively definite for stationary processes. Then, since $\langle\psi|\Sigma^<|\psi\rangle=\langle\psi|(E-H-\Sigma^a)G^<(E-H-\Sigma^r)|\psi\rangle>0$, $-i\Sigma^<$ must also be positively definite.

Due to the fact that $\Sigma^<$ and $G^<$ are related by \refeq{sigmas_comp_l} and \refeq{lesser_scatt_express}, it seems likely that $\Lambda$ should preserve positive definiteness. First of all, although it is possible that $-i\Sigma^<$ is positively definite while $-i\Sigma^<_s$ is not, this seems rather controversial, since the lead contribution to \refeq{sigmas_comp_l} can be chosen arbitrarily. Secondly, this same freedom suggests that any Greens' function $G^<$ could in principle be a solution of \refeq{eq_for_gl}, if the lead self energies were chosen appropriately. Thus, in order for $-i\Sigma^<_s$ to always be positively definite, the integral of $\Lambda(E,E',G(E))$ must be positively definite for any positively definite function $G(E)$. This must also be the case for $G(E)\sim\delta(E-E')$, leading to a positive definite $\Lambda(E,E',G)$ for any $G$.

\subsection{Assumption (iv), Charge conservation}\label{apdix_cc}

By combining \refeq{eq_for_gl} and \refeq{A_properties}, we see that 
\begin{align}\label{assump4_A_sigma_relation}
\mathrm{Tr}\,\Gamma(E)G^<(E)&=\mathrm{Tr}\,\Gamma(E)G^r(E)\Sigma^<(E)G^a(E)\\\nonumber
&=\mathrm{Tr}\,G^a(E)\Gamma(E)G^r(E)\Sigma^<(E)=\mathrm{Tr}\,A(E)\Sigma^<(E),
\end{align}
while from \refeq{negfcurrent}, the currents at the leads $p$ are given by
\begin{align}
I_p=-\frac{i}{h}\int\mathrm{d}E\,\mathrm{Tr}\,\Gamma_p(E)\left(G^<(E)-iA(E)f_p(E)\right).
\end{align}
Due to charge conservation, the total current exiting/entering the system must be zero in a stationary process. With the assistance of \refeq{sigmas_comp_r}, \refeq{sigmas_comp_l}, and \refeq{assump4_A_sigma_relation}, we get
\begin{align}
\sum_pI_p&=-\frac{i}{h}\int\mathrm{d}E\,\sum_p\mathrm{Tr}\,\Gamma_p(E)\left(G^<(E)-iA(E)f_p(E)\right)\\\nonumber
&=-\frac{i}{h}\int\mathrm{d}E\left(\mathrm{Tr}\,\left[\Gamma(E)-\Gamma_s(E)\right]G^<(E)-\mathrm{Tr}\,\left[\Sigma^<(E)-\Sigma^<_s(E)\right]A(E)\right)\\\nonumber
&=\frac{i}{h}\int\mathrm{d}E\left(\mathrm{Tr}\,\Gamma_s(E)G^<(E)-\mathrm{Tr}\,\Sigma^<_s(E)A(E)\right)=0.
\end{align}
Inserting $\Sigma^<$ from \refeq{lesser_scatt_express}, this becomes
\begin{align}
\int\mathrm{d}E\,\mathrm{Tr}\,\Gamma_s(E)G^<(E)=\int\mathrm{d}E\int\mathrm{d}E'\,\mathrm{Tr}\,\Lambda(E,E',G^<(E'))A(E),
\end{align}
which, on exchanging the integration variables on the right, becomes 
\begin{align}\label{asp4_integrated}
\int\mathrm{d}E\,\left(\mathrm{Tr}\,\Gamma_s(E)G^<(E)-\int\mathrm{d}E'\,\mathrm{Tr}\,A(E')\Lambda(E',E,G^<(E))\right)=0.
\end{align}
Following the reasoning at the end of appendix \ref{apdix_pos}, it is reasonable that any Green's function $G^<$ could in principle be a solution of \refeq{eq_for_gl}, so that \refeq{asp4_integrated} must apply also for $G^<(E)$ proportional to a delta function $\delta(E-E')$. Thus, we must have
\begin{align}
\mathrm{Tr}\,\Gamma_s(E)G=\int\mathrm{d}E'\,\mathrm{Tr}\,A(E')\Lambda(E',E,G),
\end{align}
for any positively definite and Hermitian $G$.

\subsection{Assumption (v), Detailed balance}\label{apdix_detb}

With elastic scattering, \refeq{electron_phonon} becomes instead
\begin{align}
\Sigma^<_{sij}(E)=\sum_{kl}D^{ij}_{kl}(E)G^<_{kl}(E),
\end{align}
and accordingly, $\Lambda(E,G)$ becomes
\begin{align}
\Lambda(E,G)_{ij}=\sum_{kl}D^{ij}_{kl}(E)G_{kl},
\end{align}
so that
\begin{align}
\mathrm{Tr}\,F\Lambda(E,G)=\sum_{ijkl}D^{ij}_{kl}(E)G_{kl}F_{ji}=\sum_{ijkl}D^{lk}_{ji}(E)F_{kl}G_{ji}.
\end{align}
Thus, assumption (v) would follow if we could say $D^{ij}_{kl}(E)=D^{lk}_{ji}(E)$.

In the case of electron phonon interactions, we have from \refeq{d_of_electron_phonon} and \refeq{drse_defined}
\begin{align}
D^{ij}_{kl}(E)=\frac{1}{h}\sum_{rs}U^r_{ik} U^s_{lj}\left\langle x_r(t) x_s(t)\right\rangle=\frac{1}{h}\sum_{sr}U^s_{ik} U^r_{lj}\left\langle x_s(t) x_r(t)\right\rangle =D^{lk}_{ji}(E),
\end{align}
as required. Similar expressions apply for interactions with other fields, while for disorder scattering we have from \refeq{d_of_dissorder}
\begin{align}
D^{ij}_{kl}(E)=\langle V_{ik} V_{lj}\rangle =\langle V_{lj} V_{ik}\rangle = D^{lk}_{ji}(E)
\end{align}
where $V$ is now the disorder potential. Thus, assumption (v) applies at least to these commonly used scattering models.

\section{Expansion of the spectral density}\label{apdix_detailed}

In addition to \refeq{elastic_xi_op} and \refeq{ap_defined}, we introduce the similar definitions
\begin{align}
\hat{\Xi} G(E)&=G^a(E)\Lambda(E,G(E))G^r(E),\,\,\text{and}\\
\hat{A}_q&=G^a(E)\Gamma_q(E) G^r(E).
\end{align}
Using \refeq{A_properties} we may then expand the quantity $\mathrm{Tr}\,A(E)\Gamma_q(E)$ as
\begin{align}\label{appB_first_expt}
\mathrm{Tr}\,A(E)\Gamma_q(E)&=\mathrm{Tr}\,G^r(E)\Gamma(E)G^a(E)\Gamma_q(E)=\mathrm{Tr}\,\Gamma(E)G^a(E)\Gamma_q(E)G^r(E)\\\nonumber
&=\sum_p\mathrm{Tr}\,\Gamma_p(E)\hat{A}_q(E)+\mathrm{Tr}\,\Gamma_s(E)\hat{A}_q(E).
\end{align}

With elastic scattering, assumption (iv) simplifies to
\begin{align}
\mathrm{Tr}\,\Gamma_s(E)G=\mathrm{Tr}\,A(E)\Lambda(E,G),
\end{align}
and using this we can rewrite the last term in \refeq{appB_first_expt} as
\begin{align}\label{appB_lastt_expt}
\mathrm{Tr}\,\Gamma_s(E)\hat{A}_q(E)=\mathrm{Tr}\,A(E)\Lambda(E,\hat{A}_q(E)).
\end{align}
Using \refeq{A_properties} again, we can expand \refeq{appB_lastt_expt} as
\begin{align}
\mathrm{Tr}\,\Gamma_s(E)\hat{A}_q(E)&=\mathrm{Tr}\,G^r(E)\Gamma(E)G^a(E)\Lambda(E,\hat{A}_q(E))\\\nonumber
&=\mathrm{Tr}\,\Gamma(E)G^a(E)\Lambda(E,\hat{A}_q(E))G^r(E)=\mathrm{Tr}\,\Gamma(E)\hat{\Xi}\hat{A}_q(E)\\\nonumber
&=\sum_p\mathrm{Tr}\,\Gamma_p(E)\hat{\Xi}\hat{A}_q(E)+\mathrm{Tr}\,\Gamma_s(E)\hat{\Xi}\hat{A}_q(E).
\end{align}

Continuing this process, we can in general expand
\begin{align}
\mathrm{Tr}\,\Gamma_s(E)\hat{\Xi}^n\hat{A}_q(E)&=\mathrm{Tr}\,A(E)\Lambda(E,\hat{\Xi}^n\hat{A}_q(E))=\mathrm{Tr}\,\Gamma(E)\hat{\Xi}^{n+1}\hat{A}_q(E)\\\nonumber
&=\sum_p\mathrm{Tr}\,\Gamma_p(E)\hat{\Xi}^{n+1}\hat{A}_q(E)+\mathrm{Tr}\,\Gamma_s(E)\hat{\Xi}^{n+1}\hat{A}_q(E).
\end{align}
and inserting all this back into \refeq{appB_first_expt} we find
\begin{align}
\mathrm{Tr}\,A(E)\Gamma_q(E)=\sum_{n=0}^m\sum_p\mathrm{Tr}\,\Gamma_p(E)\hat{\Xi}^n\hat{A}_q(E)+\mathrm{Tr}\,\Gamma_s(E)\hat{\Xi}^m\hat{A}_q(E).
\end{align}
Assuming that the last term goes to zero as $m\rightarrow\infty$, this becomes
\begin{align}\label{appB_general_expt}
\mathrm{Tr}\,A(E)\Gamma_q(E)=\sum_{n=0}^\infty\sum_p\mathrm{Tr}\,\Gamma_p(E)\hat{\Xi}^n\hat{A}_q(E).\\\nonumber
\end{align}

Using assumption (v) and \refeq{ap_defined}, we can rewrite the terms of \refeq{appB_general_expt} as
\begin{align}
\mathrm{Tr}\,\Gamma_p(E)\hat{\Xi}^n\hat{A}_q(E)&=\mathrm{Tr}\,\Gamma_p(E)G^a(E)\Lambda(E,\hat{\Xi}^{n-1}\hat{A}_q(E))G^r(E)\\\nonumber
&=\mathrm{Tr}\,A_p(E)\Lambda(E,\hat{\Xi}^{n-1}\hat{A}_q(E))=\mathrm{Tr}\,\hat{\Xi}^{n-1}\hat{A}_q(E)\Lambda(E,A_p(E))\\\nonumber
&=\mathrm{Tr}\,G^a(E)\Lambda(E,\hat{\Xi}^{n-2}\hat{A}_q(E))G^r(E)\Lambda(E,A_p(E))\\\nonumber
&=\mathrm{Tr}\,\Lambda(E,\hat{\Xi}^{n-2}\hat{A}_q(E))\Xi A_p(E)=\mathrm{Tr}\,\hat{\Xi}^{n-2}\hat{A}_q(E)\Lambda(E,\Xi A_p(E))\\\nonumber
&=\cdots=\mathrm{Tr}\,\hat{\Xi}^{n-k-1}\hat{A}_q(E)\Lambda(E,\Xi^k A_p(E))=\cdots\\\nonumber
&=\mathrm{Tr}\,\hat{A}_q(E)\Lambda(E,\Xi^{n-1} A_p(E))\\\nonumber
&=\mathrm{Tr}\,G^a(E)\Gamma_q(E)G^r(E)\Lambda(E,\Xi^{n-1} A_p(E))=\mathrm{Tr}\,\Gamma_q(E)\Xi^n A_p(E),
\end{align}
and we thus finally obtain
\begin{align}\label{appB_final_expt}
\mathrm{Tr}\,A(E)\Gamma_q(E)=\sum_{n=0}^\infty\sum_p\mathrm{Tr}\,\Gamma_q(E)\Xi^n A_p(E).\\\nonumber
\end{align}

\begin{acknowledgments}
We would like to acknowledge the Research Council of Norway (NANO2021 project Thelma, number 228854) for providing funding, and the Norwegian Metacenter for Computational Resources (NOTUR) for providing computational resources.
\end{acknowledgments}

\bibliography{paper1}

\end{document}